\documentclass{aa}
\usepackage[export]{adjustbox}
\usepackage{graphicx}
\usepackage{txfonts}
\begin{document}

   \title{Mirror, mirror on the outflow cavity wall}

   \subtitle{Near-infrared CO overtone disc emission of the high-mass YSO  IRAS~11101-5829\thanks{Based on observations collected at the European Southern Observatory Paranal, Chile, Programme ID 0101.C-0317(A) and 098.C-0636(A).}}

   \author{R. Fedriani\inst{1,2}
          \and A. Caratti o Garatti\inst{1,2}
          \and M. Koutoulaki\inst{1,2}
          \and R. Garcia-Lopez\inst{1,2}
          \and A. Natta\inst{1}
          \and R. Cesaroni\inst{3}
          \and R. Oudmaijer\inst{4}
          \and D. Coffey\inst{2,1}
          \and T. Ray\inst{1}
          \and B. Stecklum\inst{5}
          }

   \institute{Dublin Institute for Advanced Studies, School of Cosmic Physics, 31 Fitzwilliam Place, Dublin 2, Ireland\\
              \email{fedriani@cp.dias.ie}
          \and
             University College Dublin, School of Physics, Belfield, Dublin 4, Ireland
          \and
             INAF, Osservatorio Astrofisico di Arcetri, Largo E. Fermi 5, 50125 Firenze, Italy        
          \and
             School of Physics and Astronomy, University of Leeds, Leeds LS2 9JT, UK
          \and
             Th\"uringer Landessternwarte Tautenburg, Sternwarte 5, 07778 Tautenburg, Germany
             }

   \date{Received month day year / accepted month day year}

  \abstract
  {}
   {The inner regions of high-mass protostars are often invisible in the near-infrared, obscured by thick envelopes and discs. We aim to investigate the inner gaseous disc of IRAS~11101-5829 through scattered light from the outflow cavity walls.}
   {We observed the immediate environment of the high-mass young stellar object IRAS~11101-5829 and the closest knots of its jet, HH135-136, with the integral field unit VLT/SINFONI. We also retrieved archival data from the high-resolution long-slit spectrograph VLT/X-shooter. We analysed imaging and spectroscopic observations to discern the nature of the near-infrared CO emission.}
   {We detect the first three bandheads of the $\upsilon=2-0$ CO vibrational emission for the first time in this object. It is coincident with continuum and Br$\gamma$ emission and extends up to $\sim10\,000$\,au to the north-east and $\sim10\,000$\,au to the south-west. The line profiles have been modelled as a Keplerian rotating disc assuming a single ring in local thermodynamic equilibrium. The model output gives a temperature of $\sim3000$\,K, a CO column density of $\sim1\times10^{22}\mathrm{\,cm^{-2}}$, and a projected Keplerian velocity $\varv_\mathrm{K}\sin i_\mathrm{disc} \sim 25\mathrm{\,km\,s^{-1}}$, which is consistent with previous modelling in other high-mass protostars. In particular, the low value of $\varv_\mathrm{K}\sin i_\mathrm{disc}$ suggests that the disc is observed almost face-on, whereas the well-constrained geometry of the jet imposes that the disc must be close to edge-on. This apparent discrepancy is interpreted as the CO seen reflected in the mirror of the outflow cavity wall.}
   {From both jet geometry and disc modelling, we conclude that all the CO emission is seen through reflection by the cavity walls and not directly. This result implies that in the case of highly embedded objects, as for many high-mass protostars, line profile modelling alone might be deceptive and the observed emission could affect the derived physical and geometrical properties; in particular the inclination of the system can be incorrectly interpreted.}


   \maketitle



\section{Introduction}\label{sect:introduction}

Theoretical and observational studies support the idea that high-mass young stellar objects (HMYSOs, $M_{*}>8\,M_\odot, L_\mathrm{bol}>5\times10^3L_\odot$) might be born as a scaled-up version of their low-mass counterparts \citep[see e.g.][]{beuther2007,tan2014}. A particularly important piece of the puzzle has been the discovery of accretion discs around HMYSOs \citep[e.g.][]{patel2005,kraus2010}. Structures in Keplerian rotation have been observed in a wide range of massive protostars and they have been associated with accretion discs or toroids \citep[see e.g.][and references therein]{beltran2016}. These structures are mainly composed of gas and dust with a variety of atoms and molecules \citep[see][for a review]{henning2013}. In the sub-millimetre, millimetre, and radio regimes, we can analyse the molecules forming further away in the disc and  dust continuum \citep{cesaroni2005,cesaroni2006,cesaroni2007,motogi2019}. However, if we want to probe the inner gaseous disc within a few astronomical units from the central source, namely where accretion and ejection take place, we need to observe in the near-infrared (NIR). An excellent tracer to study the inner gaseous disc is the NIR ${}^{12}$C${}^{16}$O overtone bandhead emission (hereafter CO emission) at $2.29-2.5\mathrm{\,\mu m}$ \citep{dullemond2010}. Indeed, this emission has been observed in a number of HMYSOs \citep{scoville1983,bik2004,bik2006,davies2010,cooper2013} as well as in intermediate- and low-mass YSOs \citep[see e.\,g.][]{connelley2010,ilee2014,koutoulaki2019}. Even though the CO emission has been observed in several YSOs in multiple mass regimes, the detection rate is very low, i.e. around 20\% \citep[e.g.][]{carr1989,ishii2001,connelley2010,cooper2013}. A plausible reason for this low detection rate, at least in the case of HMYSOs, is that the CO emission seems to be sensitive to the mass accretion rate $(\dot{M}_\mathrm{acc})$ of the system; because this rate is a moderate value of $\dot{M}_\mathrm{acc}\sim10^{-5}\,M_\odot\mathrm{\,yr^{-1}}$ it best produces the most prominent CO emission \citep{ilee2018}. Therefore, adding objects with CO detection represents both a challenge and an important contribution. The CO emitting region is usually modelled as a disc in Keplerian rotation \citep[see e.g.][]{kraus2000,ilee2013} and thus can give us important constraints on the physical properties of the inner gaseous disc and the geometry of the system.

IRAS~11101-5829 (also known as G290.3745+01.6615) is a HMYSO of $L_\mathrm{bol}\sim10^4\,L_\odot$ located in the eastern Carina star-forming region and is driving the Herbig-Haro (HH) objects HH\,135/136 \citep{ogura1992,tamura1997,ogura1998}. Unlike many other HMYSOs, this object does not lie in the plane of the Milky Way and consequently has not been observed in major IR or millimetre surveys (Spitzer, Herschel, or Atlasgal), although it was observed in the Midcourse Space Experiment (MSX) and Infrared Astronomical Satellite (IRAS). The distance to this object has been associated with the distance to the open cluster Stock 13 \citep{ogura1992}, which has a photometric distance of 2.65\,kpc \citep{steppe1977}. There are {\scriptsize GAIA DR2} distance measurements for this open cluster. These measurements have large uncertainties, although they are consistent \citep[$d=2.625^{+2.52}_{-1.40}$\,kpc;][]{bailerjones2018} with the photometric distance. In this work we adopt a distance of 2.7\,kpc, as in \citet{ogura1992} and \citet{gredel2006}. The parsec-scale bipolar jet has been studied in atomic ([\ion{S}{ii}], [\ion{Fe}{ii}]) and molecular (H$_2$) tracers \citep{gredel2006} and a strong helical magnetic field has been revealed through circular polarimetry \citep{chrysostomou2007}. The geometry of the jet is well-constrained and its axis lies almost in the plane of the sky \citep[$i_\mathrm{jet}\sim5^\circ{}$;][]{ogura1998}, therefore its disc must be seen close to edge-on ($i_\mathrm{disc}\sim85^\circ{}$). Notably, no CO bandhead emission nor any evidence of circumstellar disc has been previously reported for this object.

We report on the spectro-imaging results of the first 20\,000\,au of IRAS~11101-5829. In particular, we investigate the CO emission that is observed for the first time in this object. The observations and data reduction are presented in Sect.~\ref{sect:observations}, results in Sect.~\ref{sect:results}, discussion in Sect.~\ref{sect:discussion}, and conclusions in Sect.~\ref{sect:conclusions}.



\section{Observations and data reduction}\label{sect:observations}

\subsection{Very Large Telescope/SINFONI data}

IRAS~11101-5829 was observed with the Very Large Telescope (VLT) spectrograph for integral field observations in the NIR \citep[SINFONI;][]{sinfoni2003} on 2018 June 15 (Programme ID 0101.C-0317(A)) in the $K$ band ($1.95-2.5\mu$m). The field of view (FoV) of $8'' \times 8''$ was centred on the source, with a position angle (PA) east of north (E of N) of zero degrees. Spatial sampling was $125\times250\mathrm{\,mas\,pixel^{-1}}$; the smaller sampling was carried out in the northern direction. The total exposure time was 360\,s. Spatial resolution achieved using adaptive optics + natural guide star (AO+NGS) was $0.3''-0.4''$ and spectral resolution was $\mathcal{R}\sim4000$ ($75\mathrm{\,km\,s^{-1}}$). The NGS used for the AO system was 2MASS J11121780-5846425 ($B = 14.4$, $J = 13.9$ mag and separation of $22''$ from the target). Data were reduced in the standard way, using dedicated instrument software, {\scriptsize GASGANO}, standard IRAF routines, and Python custom scripts. A wavelength accuracy of $0.11$\AA~(or $\sim1.5\mathrm{\,km\,s^{-1}}$) was achieved. Flux calibration and telluric correction were performed using the photometric standard star Hip\,053018.

\begin{figure}[ht!]
\centering
\includegraphics[width=0.49\textwidth]{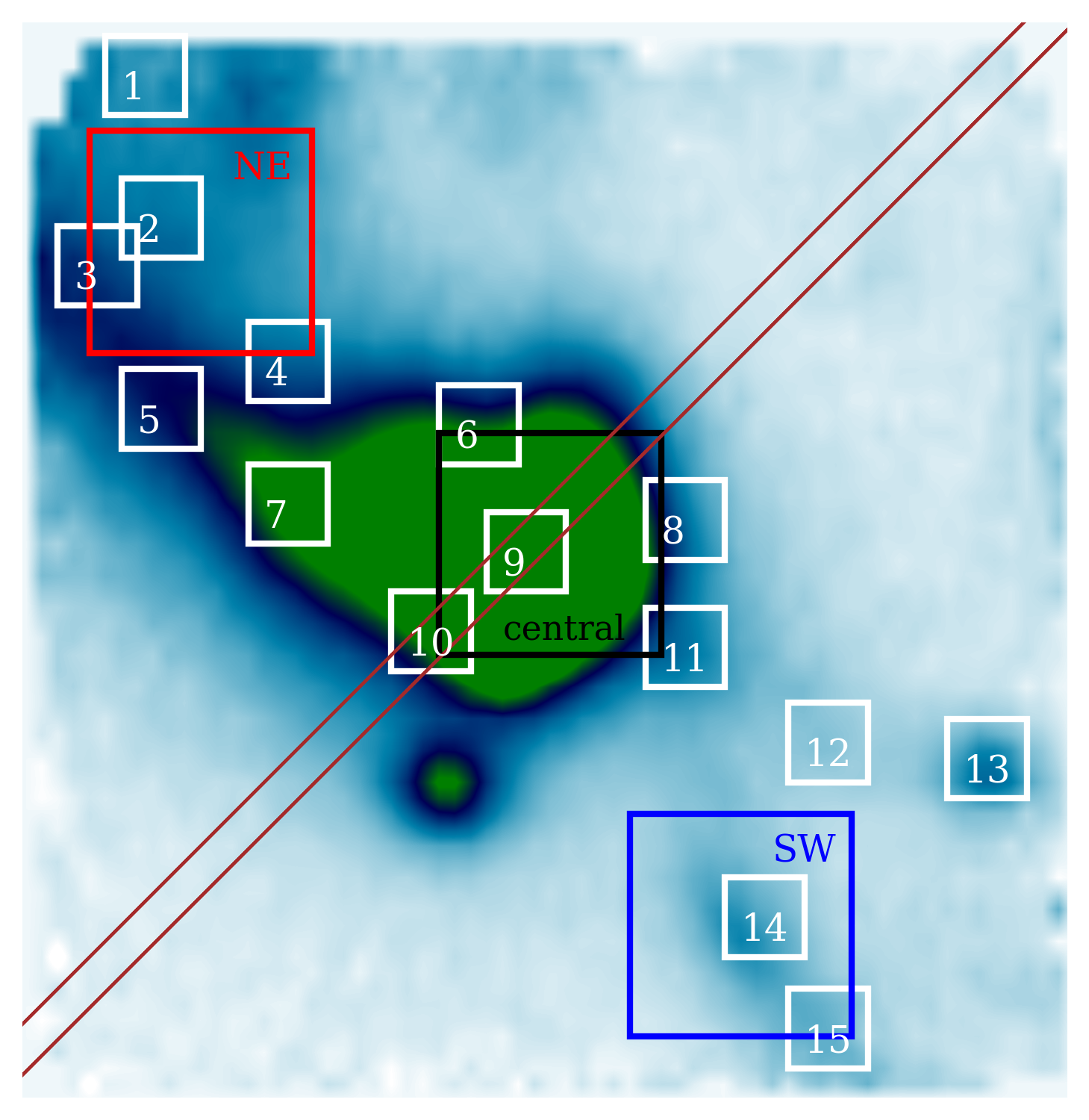}
\caption{Boxes extracted from the cube to generate the spectra shown in Figures \ref{fig:spectra_three_regions} and \ref{fig:CO_line_profile_variation} overlaid in the continuum image at $2.085\mathrm{\,\mu m}$ in arbitrary units. White boxes are $5\mathrm{\,pix}\times5\mathrm{\,pix} =0.625''\times0.625''$ whereas red, black, and blue boxes are $14\mathrm{\,pix}\times14\mathrm{\,pix} =1.75''\times1.75''$. Brown lines show the X-shooter slit position centred on the NIR peak emission.}
\label{fig:cont_boxes}
\end{figure}

Spectra at various locations of the IRAS\,11101-5829 system were extracted. For this purpose, boxes of various sizes and positions were extracted from the data cube (see Fig.~\ref{fig:cont_boxes}). Two different sets of boxes were considered. One of the sets comprises three $1.75''\times1.75''$ boxes, which are shown as red (NE), black (central), and blue (SW) rectangles in Figure~\ref{fig:cont_boxes}; the spectra are shown in Figure~\ref{fig:spectra_three_regions}. The purpose of these boxes is twofold: measure the line fluxes at each position and discern whether or not there is large-scale variation in the line profiles along the red-shifted and blue-shifted cavity walls and on source (NE, SW, and central box, respectively). The second set consists of a total of 15 boxes, each $0.625''\times0.625''$ in size, distributed along the system. The purpose of these boxes is to discern if there is any line profile variation with distance and/or location from the source (see Fig.~\ref{fig:CO_line_profile_variation}).

\subsection{Very Large Telescope/X-shooter archival data}

High-resolution long-slit spectra using the VLT/X-shooter \citep{xshooter2011} were also retrieved (Programme ID 098.C-0636(A), observed on 2017 January 31). From the full data set, we only used a small portion of the NIR arm and thus only technical details for this arm are given. The $0.4'' \times 11''$ long slit was positioned at the brightest point of the NIR nebula with a PA of $-45^\circ{}$ E of N, perpendicular to the jet (see Fig.~\ref{fig:cont_boxes}). Total exposure time was 560\,s. The spectral resolution achieved was $\mathcal{R}\sim7000$ ($\sim43\mathrm{\,km\,s^{-1}}$) and the seeing-limited spatial resolution was $\sim1.0''$. A wavelength calibration accuracy of $0.3$\AA~(or $\sim4.2\mathrm{\,km\,s^{-1}}$) was achieved. We retrieved the pipeline-calibrated spectra from the ESO Archive Science Portal. The spectrum was corrected for telluric absorption features using the telluric standard Hip\,058859. The X-shooter and SINFONI CO bandhead profiles, extracted from the same region, are consistent with each other (see Fig.~\ref{fig:CO_spectrum_xs_sin}).



\section{Results}\label{sect:results}

\begin{figure*}[ht!]
\centering
\includegraphics[width=1.0\textwidth]{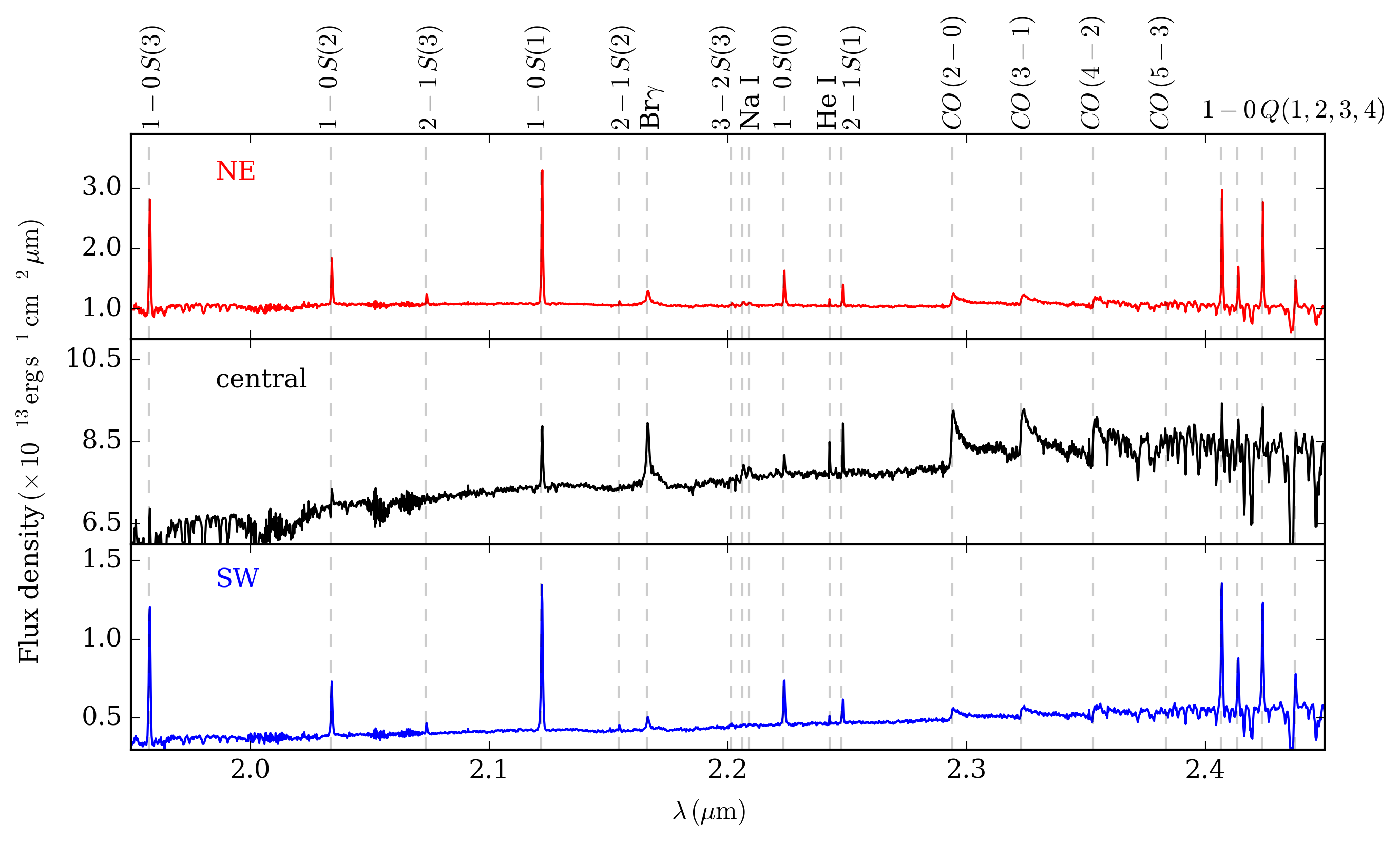}
\caption{VLT/SINFONI spectra in three different regions (see Fig.~\ref{fig:cont_boxes} for the boxes correspondence). The detected lines are indicated on the top.}
\label{fig:spectra_three_regions}
\end{figure*}

\subsection{Spectra extracted along the IRAS\,11101-5829 system}\label{sect:emission_maps}

The main advantage of integral field unit observations is that we have imaging and spectra for each single pixel in the image, providing us with spectral and spatial information of the system. In this section, we first present the spectra extracted at various locations and then we show the emission maps of the central source and its immediate environment.

Our data show a plethora of prominent emission lines, namely H$_2$, Br$\gamma$, \ion{He}{i}, \ion{Na}{i}, and CO (see Fig.~\ref{fig:spectra_three_regions} and Table~\ref{tab:observed_lines_IRAS11101}). In this section, we focus on the results related with the CO emission and use the H$_2$ lines only to determine the jet structure and kinematics (see Sect.~\ref{sect:H2_geometry}). For the first time, the NIR CO bandhead emission at $2.29-2.4\mathrm{\,\mu m}$, associated with active accreting discs, is observed in this source. In particular, the first three bandheads and a hint of the fourth ($\upsilon=2-0, \upsilon=3-1, \upsilon=4-2, \upsilon=5-3$ transitions) are detected in both X-Shooter and SINFONI spectra. In the case of SINFONI spectra, the emission is more intense at the centre of the nebula (Fig.~\ref{fig:spectra_three_regions} central panel, see also Fig.~\ref{fig:flux_density_maps}), but it is also observed in the north-east (NE) and south-west (SW) outflow cavity walls extending more than 10\,000\,au (Fig.~\ref{fig:spectra_three_regions} top and bottom panels).

Similarly, the Br$\gamma$ at $2.1662\mathrm{\,\mu m}$, \ion{Na}{i} doublet at $2.2062/2.2089\mathrm{\,\mu m}$, and \ion{He}{i} at $2.2437\mathrm{\,\mu m}$ lines show an analogous behaviour, which is more intense at the central brightest region and weaker in NE and SW. Interestingly, all these lines are expected to form in the disc or in its proximity \citep{lorenzetti2011}. However, the large spatial extent (thousands of au) and its spatial coincidence with the continuum emission from the central source, tracing the outflow cavity walls, indicate that this emission is reflected. The line profile of the CO bandheads does not change significantly along the emitting area (see Fig.~\ref{fig:cont_boxes} and Fig.~\ref{fig:CO_line_profile_variation}) suggesting that the geometry of the outflow cavity walls does not change much \citep[see e.g.][]{davies2010}.

\begin{table*}
\caption{Observed emission lines on IRAS\,11101-5829. The fluxes were measured towards the three regions indicated in Figure~\ref{fig:cont_boxes}.}             
\label{tab:observed_lines_IRAS11101}      
\centering          
\begin{tabular}{c c c c c c}    
\hline\hline       
\noalign{\smallskip}
Species & Transition & $\lambda_\mathrm{vac}$ & \multicolumn{3}{c}{Flux}  \\ 
  & &($\mu\mathrm{m}$) &  \multicolumn{3}{c}{($10^{-14}\mathrm{\,erg\,cm^{-2}\,s^{-1}}$)} \\
  & & & NE region & SW region & central region\\
\noalign{\smallskip}
\hline              
\noalign{\smallskip}
  H$_2$ & $1-0$\,S(3) & 1.95755 & $5.10\pm0.12$ & $2.69\pm0.05$ & $\cdots$ \\
  H$_2$ & $1-0$\,S(2) & 2.03375 & $1.78\pm0.06$ & $0.89\pm0.02$ & $\cdots$ \\
  H$_2$ & $2-1$\,S(3) & 2.07351 & $0.47\pm0.04$ & $0.18\pm0.02$ & $\cdots$ \\
  H$_2$ & $1-0$\,S(1) & 2.12182 & $5.19\pm0.02$ & $2.71\pm0.01$ & $3.87\pm0.14$ \\
  H$_2$ & $2-1$\,S(2) & 2.15422 & $0.17\pm0.02$ & $0.15\pm0.01$ & $\cdots$ \\
  \ion{H}{i} & Br$\gamma$ & 2.16612 & $2.94\pm0.37$ & $0.77\pm0.13$ & $17.2\pm0.31$ \\
  H$_2$ & $3-2$\,S(3) & 2.20139 & $0.22\pm0.05$ & $0.14\pm0.03$ & $\cdots$ \\   
  \ion{Na}{i} & ${}^2S_{3/2}-{}^2Po_{1/2}$ & 2.20624 & $0.37\pm0.05$ & $\cdots$ & $1.50\pm0.02$ \\
  \ion{Na}{i} & ${}^2S_{1/2}-{}^2Po_{1/2}$ & 2.20897 & $0.44\pm0.07$ & $\cdots$ & $1.42\pm0.02$ \\
  H$_2$ & $1-0$\,S(0) & 2.22330 & $1.24\pm0.02$ & $0.84\pm0.01$ & $1.34\pm0.17$ \\
  \ion{He}{i} & ${}^1S_1-{}^1Po_0$& 2.24373 & $0.14\pm0.01$ & $0.07\pm0.01$ & $1.03\pm0.06$ \\
  H$_2$ & $2-1$\,S(1) & 2.24772 & $0.59\pm0.01$ & $0.32\pm0.01$ & $1.24\pm0.04$ \\
  CO\tablefootmark{a} & $\upsilon=2-0$ & 2.29320 & $4.48\pm0.45$ & $2.47\pm0.25$ & $41.0\pm4.10$ \\
  CO\tablefootmark{a} & $\upsilon=3-1$ & 2.32320 & $4.09\pm0.41$ & $3.21\pm0.32$ & $46.5\pm0.47$ \\
  CO\tablefootmark{a} & $\upsilon=4-2$ & 2.35320 & $2.55\pm0.26$\tablefootmark{b} & $3.60\pm0.36$\tablefootmark{b} & $41.4\pm4.14$\tablefootmark{b} \\
  CO\tablefootmark{c} & $\upsilon=5-3$ & 2.38350 & $\cdots$ & $\cdots$ & $\cdots$ \\

  H$_2$ & $1-0\,Q(1)$ & 2.40659 & $4.30\pm0.15$ & $2.49\pm0.09$ & $\cdots$ \\
  H$_2$ & $1-0\,Q(2)$ & 2.41343 & $1.61\pm0.16$ & $1.01\pm0.08$ & $\cdots$ \\
  H$_2$ & $1-0\,Q(3)$ & 2.42372 & $3.67\pm0.14$ & $2.04\pm0.09$ & $\cdots$ \\
  H$_2$ & $1-0\,Q(4)$ & 2.43749 & $1.15\pm0.16$ & $0.64\pm0.08$ & $\cdots$ \\
\noalign{\smallskip}
\hline                  
\end{tabular}
\tablefoot{$^a$ Since no Gaussian profile could be fitted, the flux was obtained integrating over the curve in the specific wavelength and the error considered to be $10\%$. $^b$ Affected by poor atmospheric transmission and telluric subtraction. $^{c}$ Emission too faint to measure the flux.}
\end{table*}

\subsection{Emission maps of the first 10\,000\,au of IRAS\,11101-5829}
\begin{figure*}[ht!]
\centering
\includegraphics[width=0.45\textwidth]{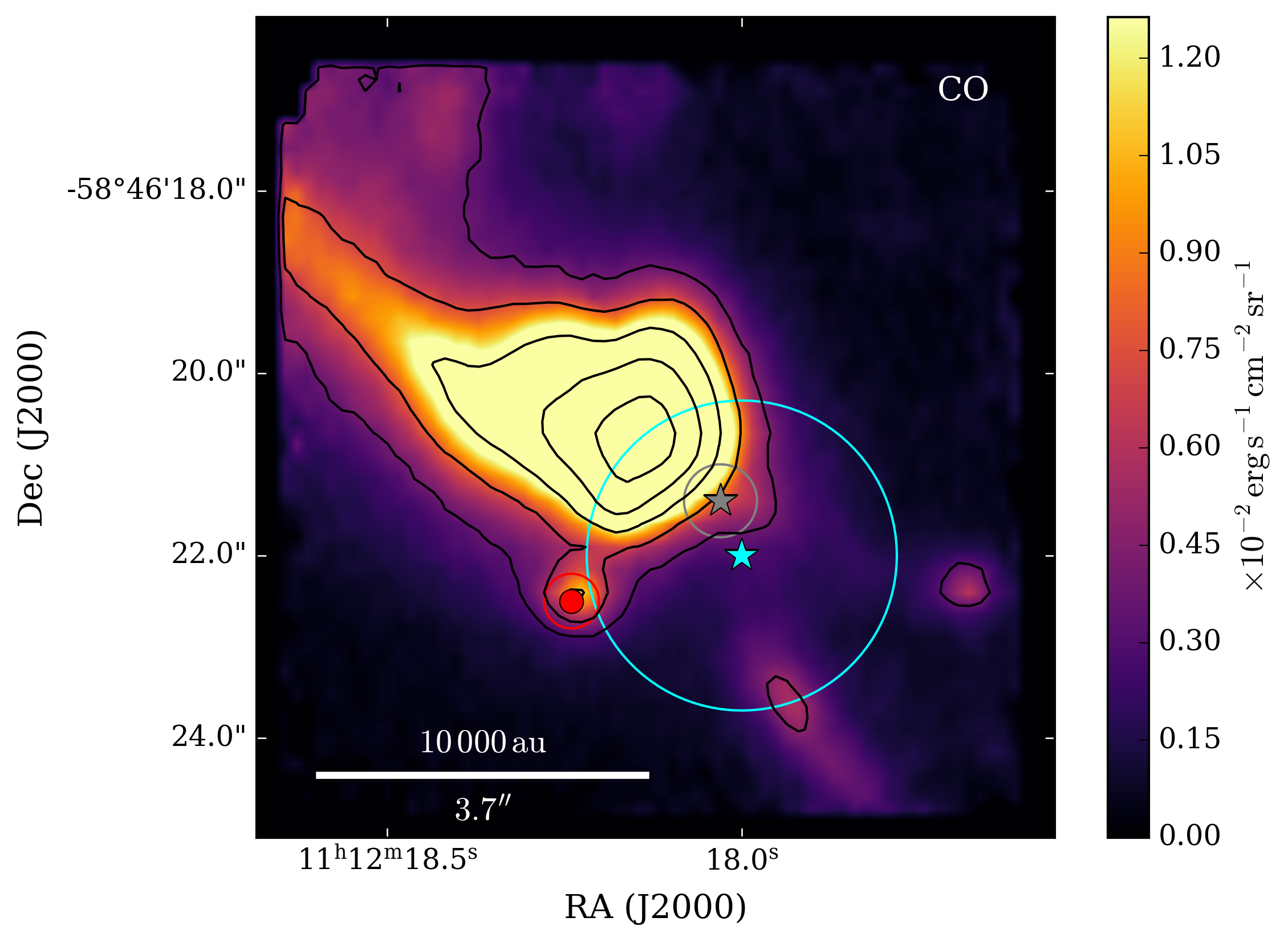}
\includegraphics[width=0.45\textwidth]{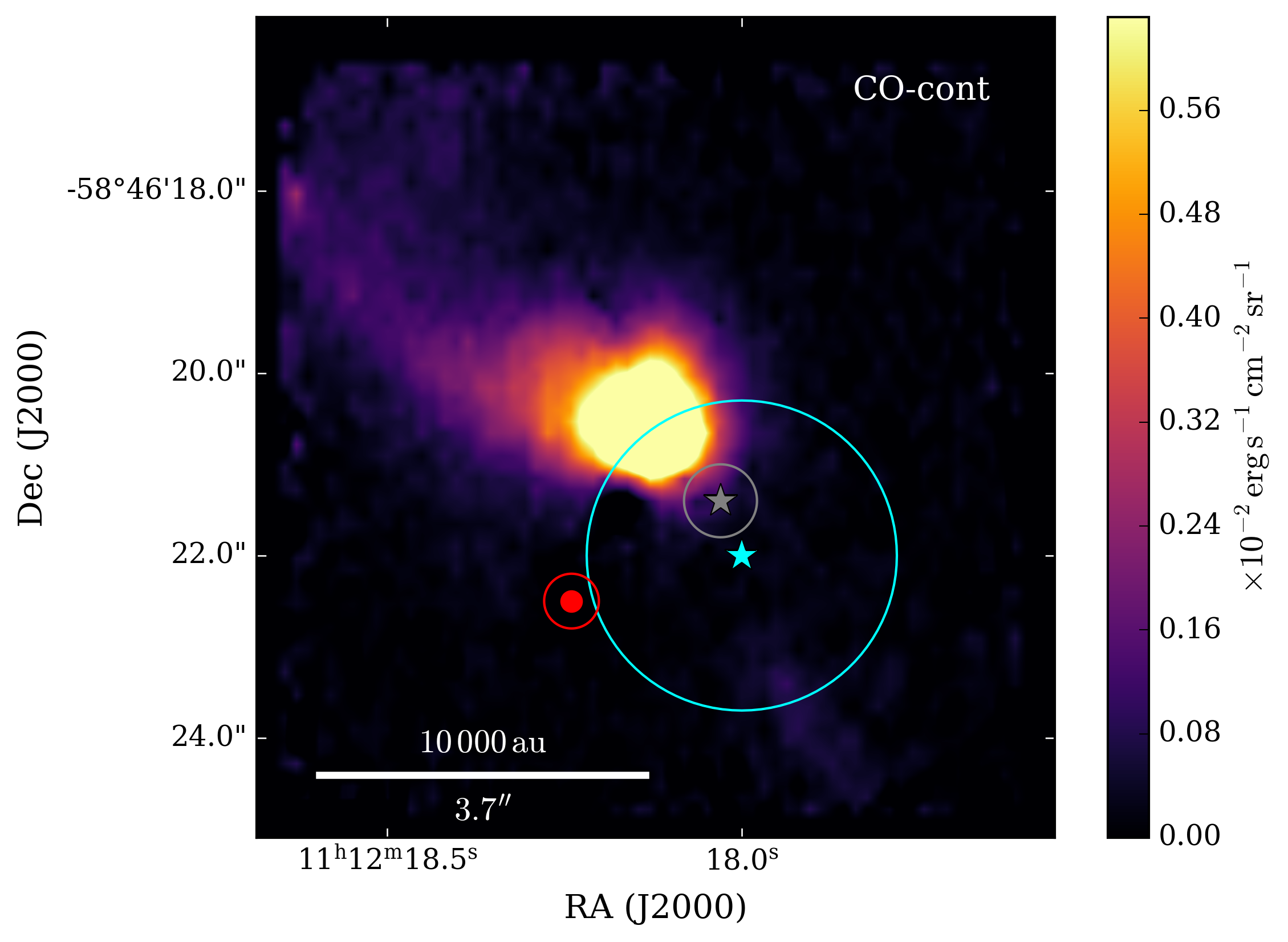}
\includegraphics[width=0.45\textwidth]{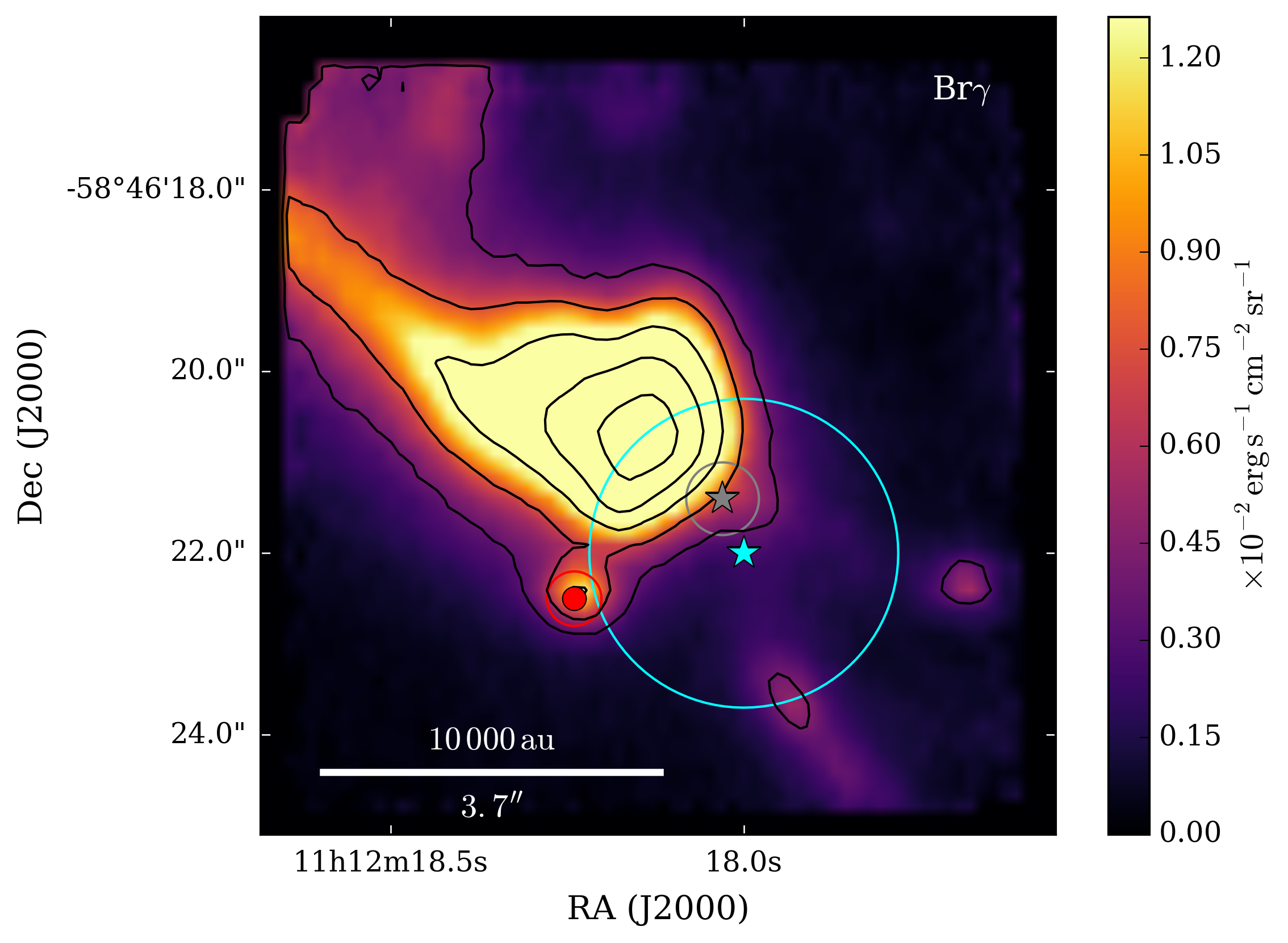}
\includegraphics[width=0.45\textwidth]{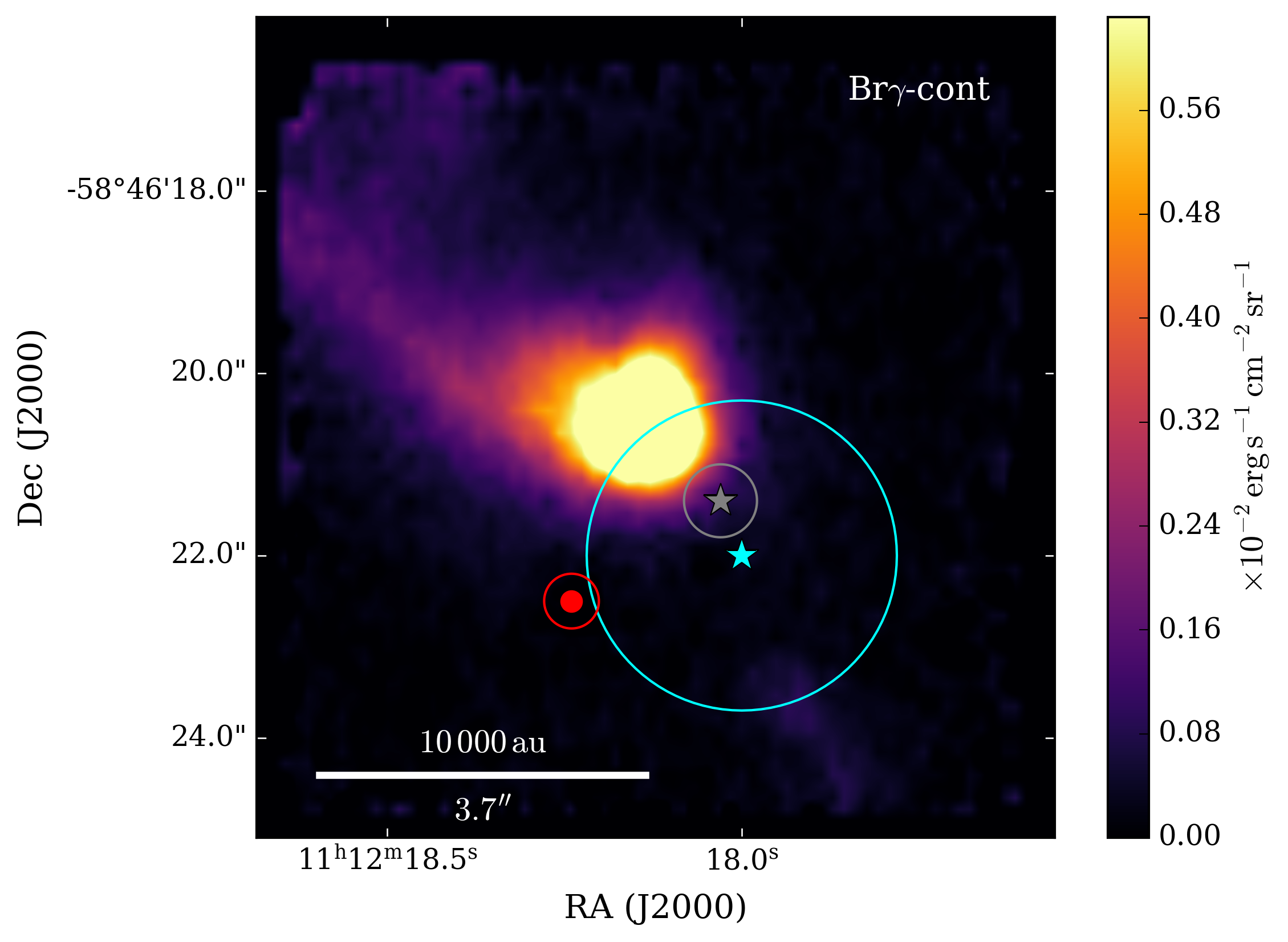}
\includegraphics[width=0.45\textwidth]{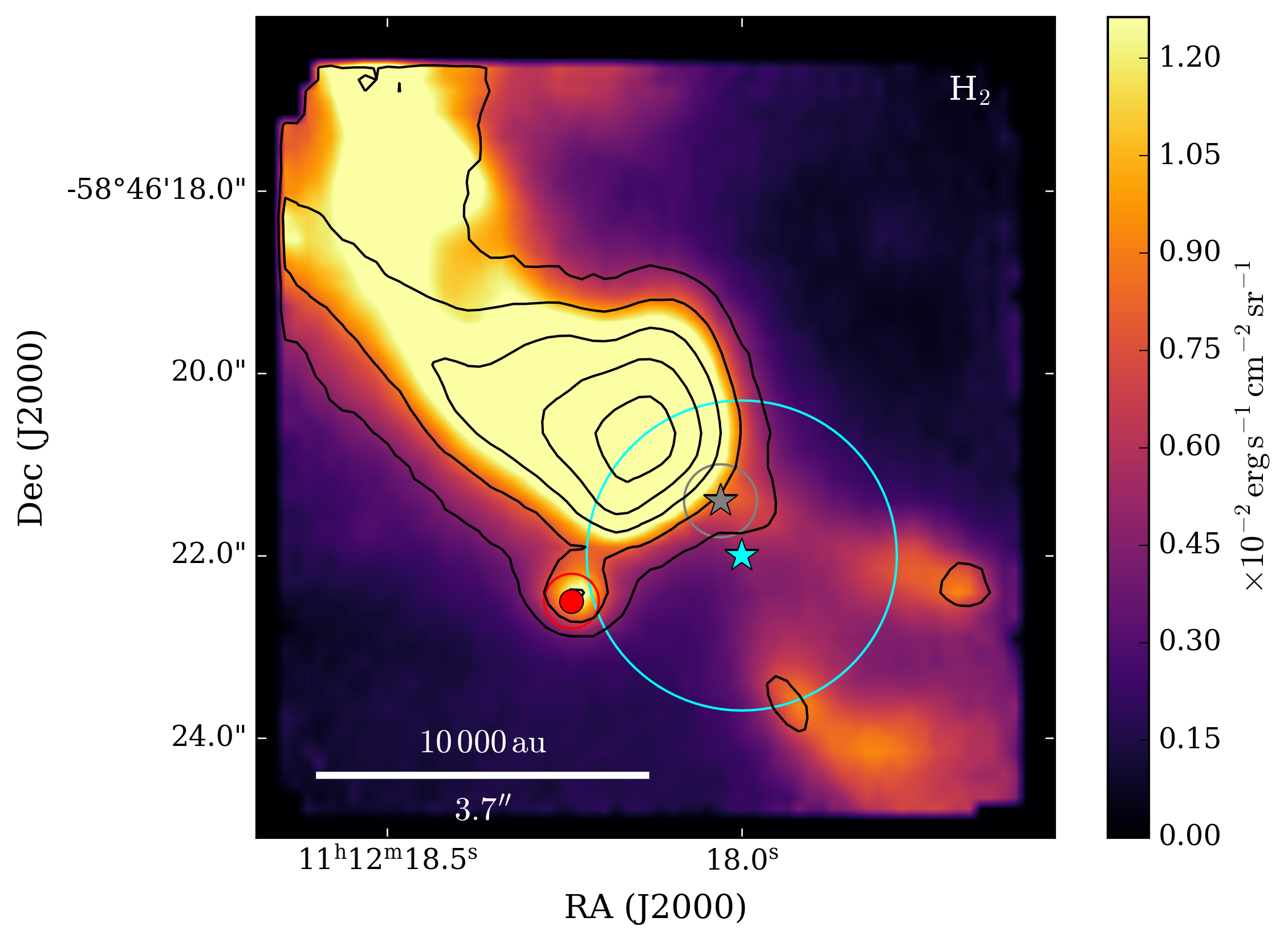}
\includegraphics[width=0.45\textwidth]{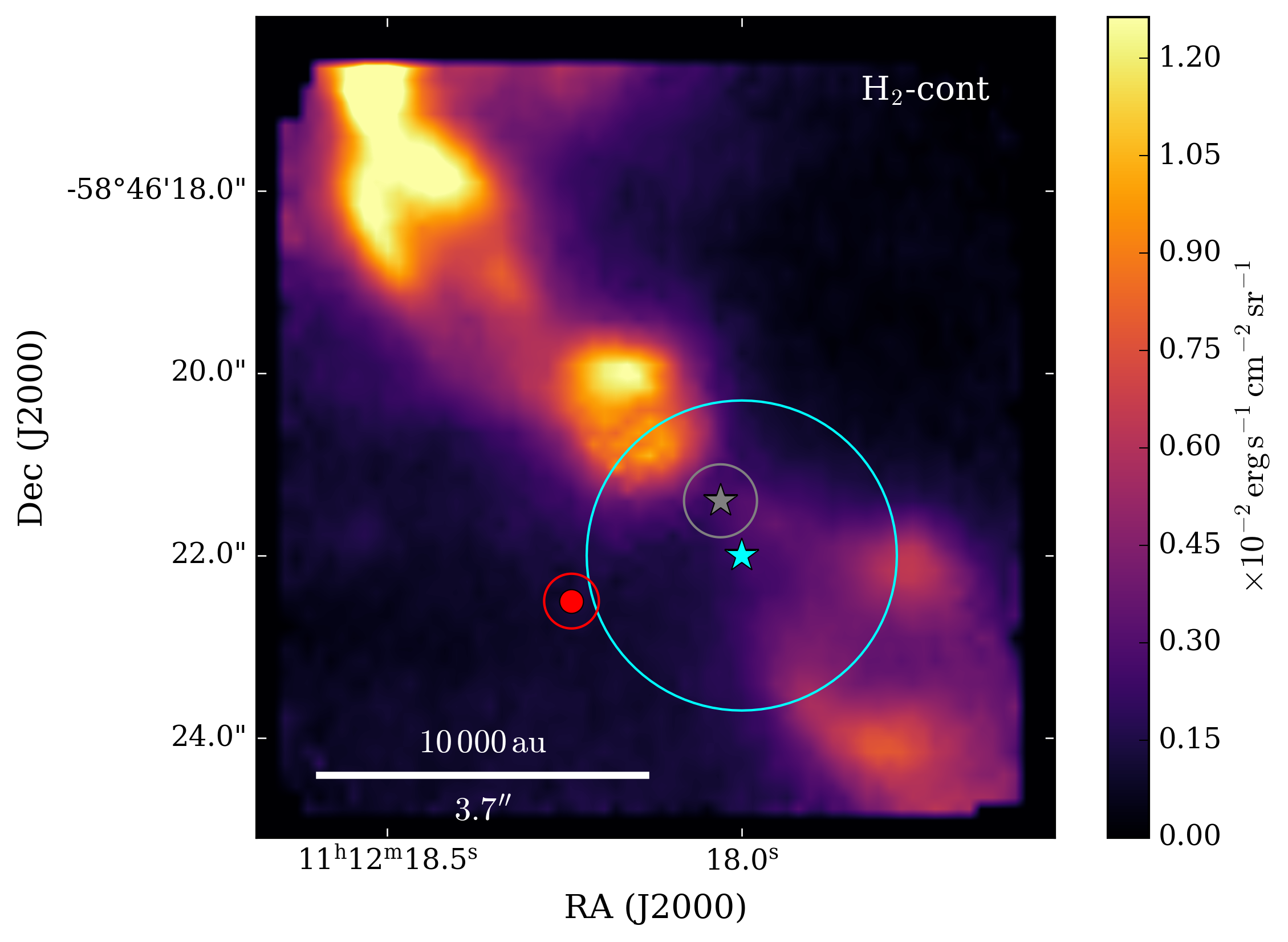}
\caption{Emission maps from VLT/SINFONI for CO (top), Br$\gamma$ (middle), and H$_2$ (bottom), integrated over the peak of each line ($\Delta\lambda\sim0.00024\,\mu$m). The black contour lines in the left panels represent the continuum emission at $2.08559\mathrm{\,\mu m}$, the levels are $(5,10,20,40,80,160)\times\sigma$ where $\sigma\sim0.1\mathrm{\,erg\,s^{-1}\,cm^{-2}\,\mu m^{-1}\,sr^{-1}}$. The right panels represent the continuum-subtracted emission maps, where the continuum subtraction was performed by considering the continuum left (blue-shifted) and right (red-shifted) of each emission line, with the exception of the CO, where only the left (blue-shifted) part was considered. The red circle represents the star in the {\scriptsize GAIA DR2} catalogue (Source ID: 5339406246100053888, RA(J2000) = 11h12m18.24s, Dec(J2000) = -58d46m22.5s) used for accurate astrometry of our cubes (the circle around it represents the uncertainty $\sim0.3''$). The cyan star represents the suggested position of the star given by \citet{tamura1997}; the circle around it represents the uncertainty $\sim1.7''$. The grey star represents the position of the central source as given by our kinematic study (see Sect.~\ref{sect:H2_geometry}). In all panels north is up and east is left.}
\label{fig:flux_density_maps}
\end{figure*}

Our SINFONI FoV covers $21\,600\mathrm{\,au}\times21\,600\mathrm{\,au}$ at a distance of $2.7$\,kpc. The top left panel of Figure~\ref{fig:flux_density_maps} shows the emission map of the first bandhead CO\,$(\upsilon=1-0)$ line at $2.2932\mathrm{\,\mu m}$; the contribution of the continuum at $2.0855\mathrm{\,\mu m}$ is represented in black contours. The continuum emission is very bright at the centre of the image and extends smoothly to the NE and towards the SW, where it becomes weaker. The continuum is tracing the emission of the central engine and its circumstellar environment being the extended emission reflected light. The top right panel of Figure~\ref{fig:flux_density_maps} shows the continuum-subtracted emission map of the CO\,$(\upsilon=1-0)$ line extending more than 10\,000\,au, mimicking both the shape and distribution of that of the continuum. This indicates that the CO emission, which comes from the circumstellar disc, is reflected in the outflow cavity walls (as for the continuum), and does not represent real extended emission directly observed from the disc.

In the middle panels, the emission maps of the Br$\gamma$ line (line+continuum and continuum-subtracted, left and right, respectively) are presented. It is clear that the morphology of both CO and Br$\gamma$ emitting regions is remarkably similar, suggesting that both are reflected in the outflow cavity walls. This idea is strengthened by the fact that these transitions have different excitation energies and trace different gas  conditions. Indeed, it should be noted that CO and Br$\gamma$ emission should come from different regions/layers of the disc \citep[see e.g.][]{dullemond2010}. In addition, the Br$\gamma$ could also be emitted from the base of a wind \citep[see e.g.][]{tambovtseva2016} or from a collimated jet close to the disc~\citep{caratti2016,fedriani2019}.

In contrast, the continuum-subtracted H$_2$ emission map, corresponding to the $1-0$\,S(1) transition at $2.1218\mathrm{\,\mu m}$, clearly delineates the molecular jet (bottom panels of Fig.~\ref{fig:flux_density_maps}), which has an orientation of NE to SW \citep[see Fig.~7 of][to see the full extent of the molecular jet]{gredel2006}. It should be noted that some H$_2$ emission might also represent scattered light in the cavity walls. Towards the NE (which corresponds to the red-shifted lobe, see Fig.~\ref{fig:H2_vel_map}) there is an evident bow-shock structure. Similarly, towards the SW (which corresponds to the blue-shifted lobe) there is also a hint of bow-shock-like structure.

\subsection{Line profile variation}

\begin{figure}[ht!]
\centering
CO
\includegraphics[width=0.49\textwidth]{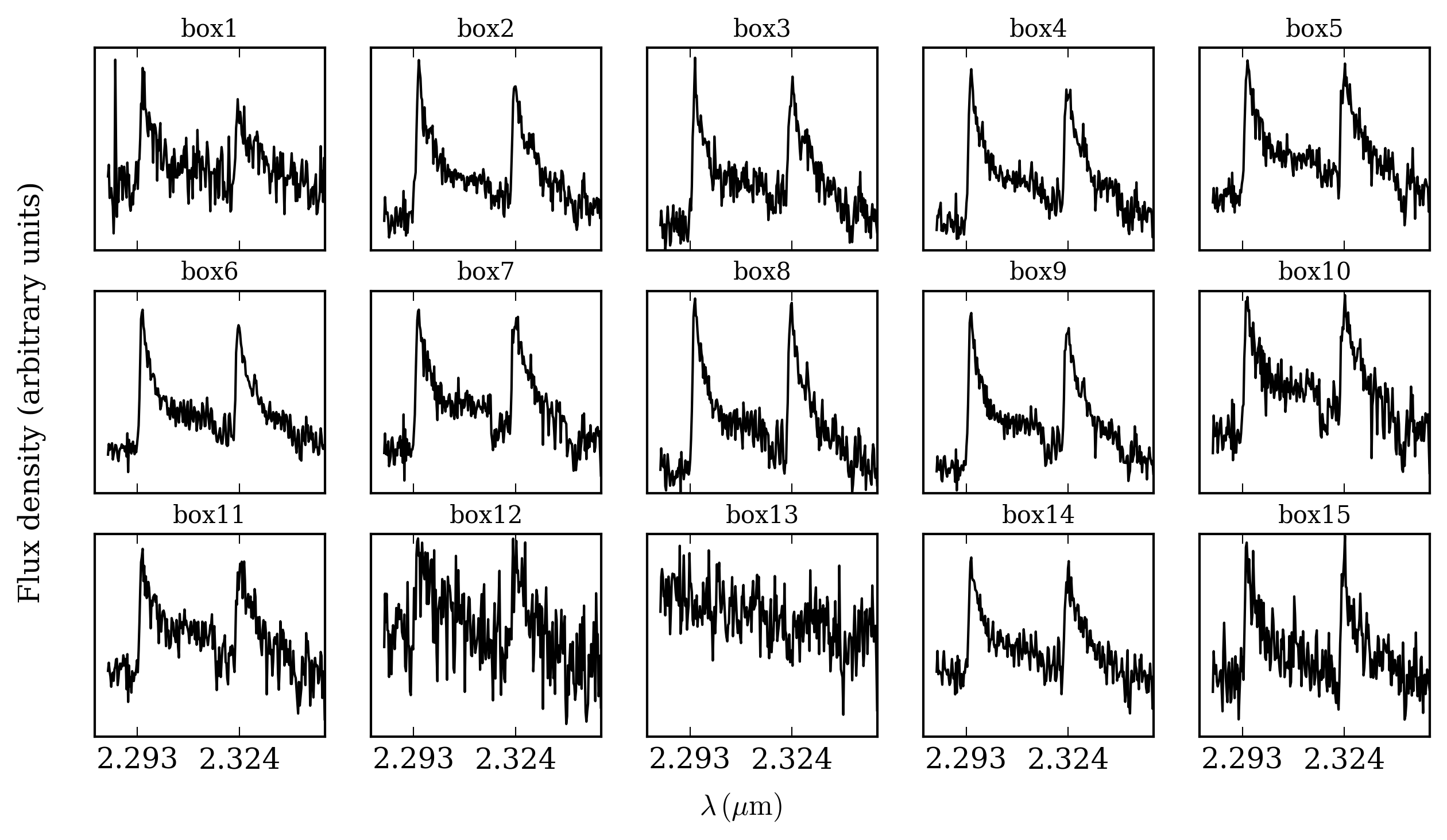}
Br$\gamma$
\includegraphics[width=0.49\textwidth]{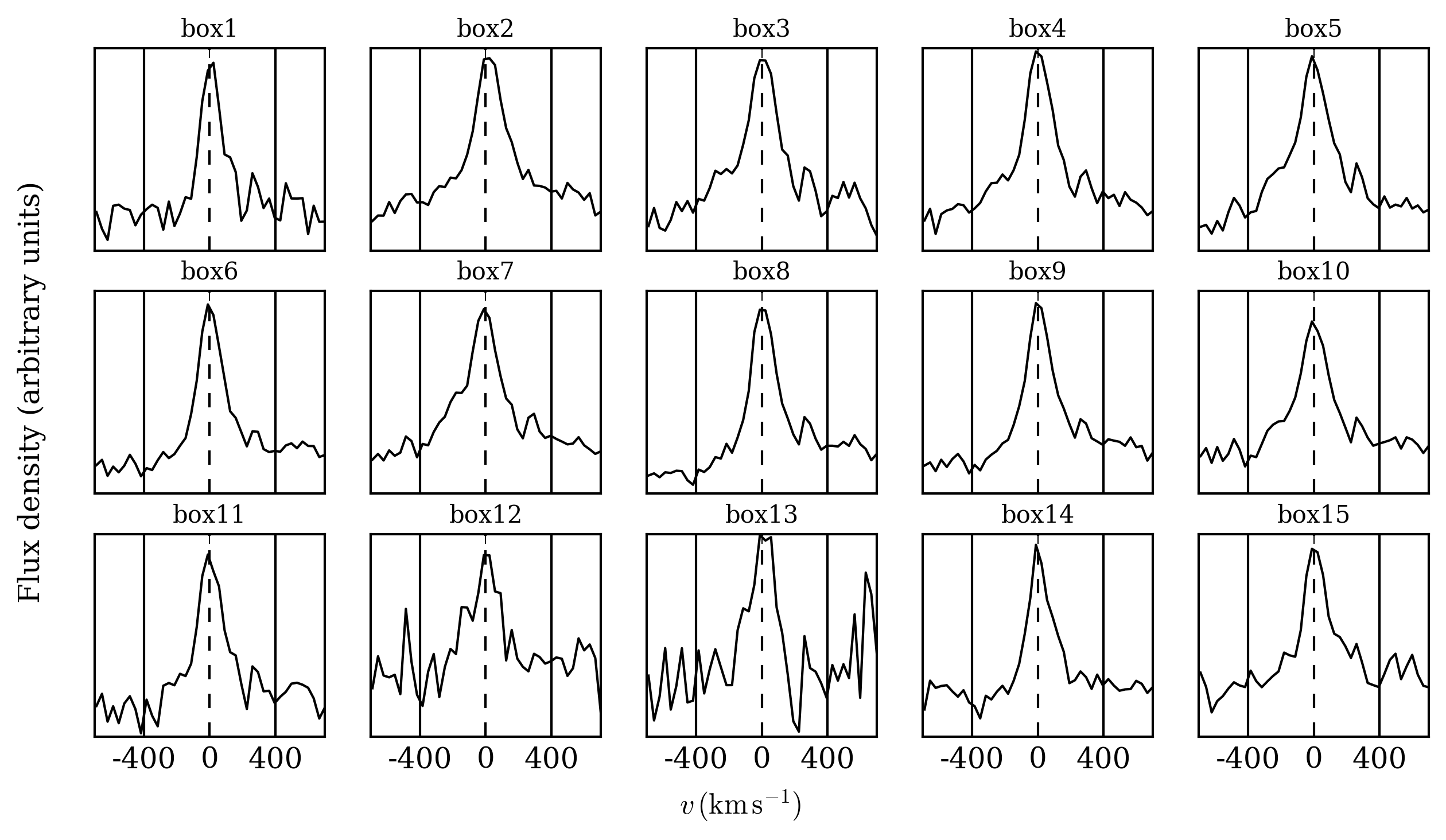}
H$_2$
\includegraphics[width=0.49\textwidth]{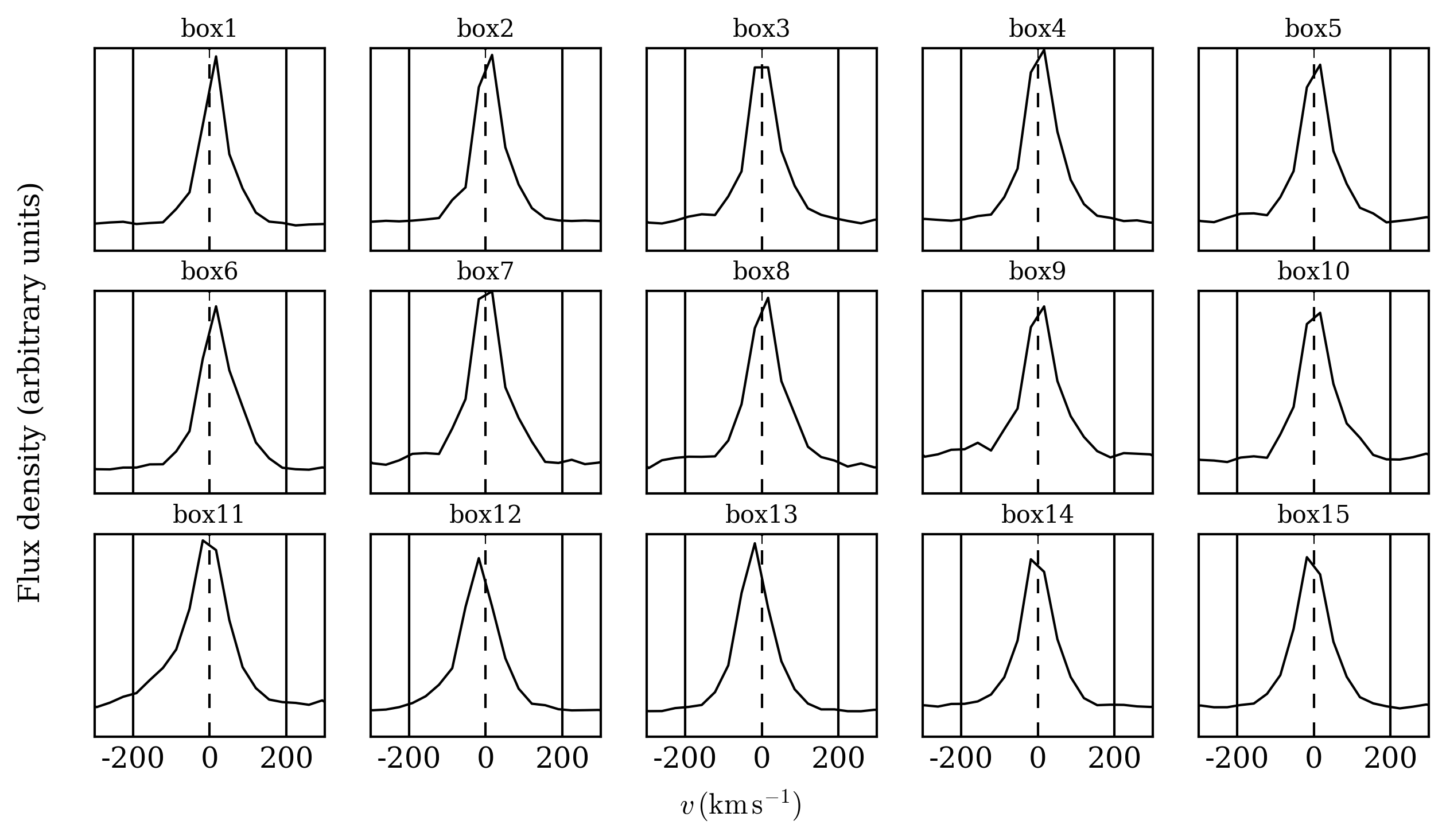}
\caption{\textit{Top panel:} Spectra from VLT/SINFONI showing the profile variation of the two firsts CO bandhead lines along the cube. \textit{Middle panel:} Same as top panel for the Br$\gamma$ line at $2.16\,\mu$m. \textit{Bottom panel:} Same as top panel for the H$_2$ line $1-0$\,S(1) at $2.12\,\mu$m. The box numbering corresponds to that given in Figure~\ref{fig:cont_boxes}. Velocities are given with respect to the LSR.}
\label{fig:CO_line_profile_variation}
\end{figure}

The line profile variations of the first two CO bandheads, the $1-0$\,S(1), Br$\gamma,$ and the H$_2$ extracted from the 15 selected regions in SINFONI FoV (see Fig.~\ref{fig:cont_boxes}) are shown in the top, middle, and bottom panels of Figure~\ref{fig:CO_line_profile_variation}, respectively.

The CO emission is detected in all the small white boxes with the exception of box\,13. Notably, there is no significant change in the line profile, therefore there is no relevant variation in the geometry of the outflow cavity walls. The same happens for the Br$\gamma$ line profiles which show radial velocity peaks around 0\,km\,s$^{-1}$. This indicates that the bulk of the Br$\gamma$ emission is seen in scattered light. There is however an extra blue-shifted component detected in some of the line profiles (e.g. boxes 3, 5, 7, and 10). This blue-shifted component might possibly trace a wind. Finally, there is a notable change in the line profile of the molecular hydrogen line $1-0$\,S(1) (Fig.~\ref{fig:CO_line_profile_variation} bottom panel). In this figure, we can clearly see that both the FWZI and the line peak vary, depending on where the spectrum was extracted. On the one hand, the FWZI varies from $\sim200$ to $400\mathrm{\,km\,s^{-1}}$. On the other hand, the difference between the red- and blue-shifted lobes is also evident. Namely, the boxes located towards the NE (i.e. boxes $1-10$) show line profiles peaking at red-shifted radial velocities, whereas those located towards the SW (i.e. boxes $11-15$) show lines profiles peaking at blue-shifted radial velocities delineating the jet (see Section~\ref{sect:H2_geometry} and Figure~\ref{fig:H2_vel_map} for more details on the velocity structure of the jet).

\subsection{H$_2$ jet kinematics to probe the geometry of the system}\label{sect:H2_geometry}

Twelve H$_2$ lines are detected (see Table~\ref{tab:observed_lines_IRAS11101}). The brightest line corresponds to the $1-0$\,S(1) transition exceeding a signal-to-noise ratio of 100. This emission is a clear tracer of jet shocked material.

\begin{figure}[ht!]
\centering
\includegraphics[width=0.49\textwidth]{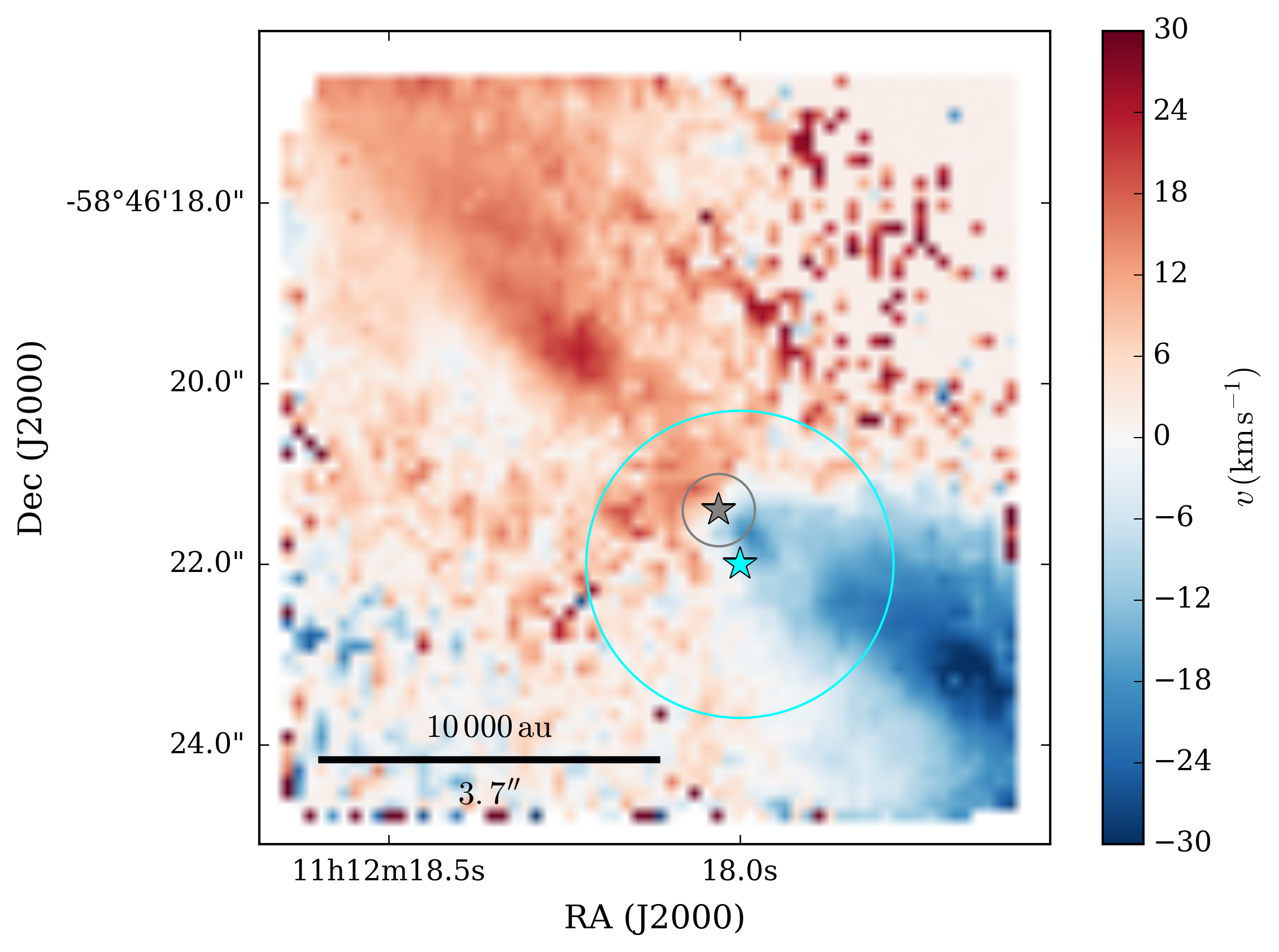}
\caption{Velocity map (LSR) of molecular hydrogen obtained from the $1-0$\,S(1) line at $2.12\,\mu$m. The cyan star denotes the position of the star given by \citet{tamura1997}; the circle around it represents the uncertainty $\sim1.7''$. The grey star indicates our suggested position of the central engine; the circle around it represents the uncertainty $\sim0.4''$.}
\label{fig:H2_vel_map}
\end{figure}

Figure~\ref{fig:H2_vel_map} shows the  $1-0$\,S(1) velocity map tracing the protostellar jet of IRAS\,11101-5829. To construct the map, we only considered pixels with signal-to-noise ratios greater than 10 in the H$_2$ line. The radial velocities shown in Figure~\ref{fig:H2_vel_map} are with respect to the local standard of rest (LSR) and corrected for the velocity of the parent cloud \citep[$\varv_\mathrm{LSR}^\mathrm{cloud}=-24\mathrm{\,km\,s^{-1}}$,][]{walsh1997}. There is a clear distinction between the red-shifted lobe (towards the NE) and the blue-shifted lobe (towards the SW). In the red-shifted lobe the radial velocities measure up to $30\mathrm{\,km\,s^{-1}}$, whereas in the blue-shifted lobe they reach up to $-30\mathrm{\,km\,s^{-1}}$.

\citet{hartigan1987} demonstrated that the full width at zero intensity (FWZI) of emission lines tracing jet shocked material is similar to the shock velocity. From our data, we measured the FWZI of the $1-0$\,S(1) line at various locations obtaining a shock velocity $(\varv_\mathrm{tot})$ of $\sim200-400\mathrm{\,km\,s^{-1}}$ (see Fig.~\ref{fig:CO_line_profile_variation}, bottom panel). The radial velocity $(\varv_\mathrm{rad})$ was also measured to be $\sim15-30\mathrm{\,km\,s^{-1}}$ (see Fig.~\ref{fig:H2_vel_map}). Hence, knowing $(\varv_\mathrm{tot})$ and $(\varv_\mathrm{rad})$, the inclination angle of the jet can be estimated as $i_\mathrm{jet}=\arcsin\left(\frac{\varv_\mathrm{rad}}{\varv_\mathrm{tot}}\right)$. We obtained a range of values of $i_\mathrm{jet}\sim2.1^\circ{}-8.6^\circ{}$ at a distance of $1^{\prime\prime}-4^{\prime\prime}$ ($2700-10\,800$\,au) from the central source. This result is consistent with \citet{ogura1998}, who obtained $i_\mathrm{jet}\sim5^\circ{}$ at a distance of $20^{\prime\prime}-60^{\prime\prime}$ ($54\,000-162\,000$\,au), utilising optical atomic lines (H$\alpha$\,$\lambda\lambda6562$ and [\ion{N}{ii}]\,$\lambda\lambda6584$).

As the jet is orthogonal to the disc and its precession is modest (see Fig.~\ref{fig:flux_density_maps} and Figs. 6 and 7 in \citeauthor{gredel2006} \citeyear{gredel2006}), its geometry allows us to infer that of the disc, and vice versa. As just shown, HH\,135/136 is nearly in the plane of the sky and therefore the IRAS\,11101-5829 disc must be near to edge-on; see for example the case of HH\,30 or HH\,111,
where the jets lie also almost in the plane of the sky \citep[][]{ray1996,reipurth1997}.

\subsection{Physical conditions of the NIR CO bandhead overtone emitting region}\label{sect:co_modelling}

We developed a local thermodynamic equilibrium (LTE) model \citep[see][]{koutoulaki2019} to explain the CO emission. The details of the model are given in the Appendix\,\ref{sect:appendix_CO}, and the main results are reported below. We assumed a single ring in LTE at temperature $T$, CO column density $N(\mathrm{CO})$, turbulence velocity of the molecule $\Delta \varv$, and projected Keplerian velocity $\varv_\mathrm{K}\sin i_\mathrm{disc}$, where $i_\mathrm{disc}$ is the inclination of the plane of the disc with respect to the plane of the sky. That is $i_\mathrm{disc}=0^\circ{}$ is face-on and $i_\mathrm{disc}=90^\circ{}$ is edge-on; we note that the latter case corresponds to the jet laying in the plane of the sky, i.e. $i_\mathrm{jet}=0^\circ{}$. We modelled SINFONI and X-shooter spectra, reproducing both spectra with very similar conditions. However, given the higher spectral resolution of X-shooter, we only considered this spectrum for our modelling. We subtracted the contribution of the continuum by fitting a straight line to the left part of the first bandhead and extrapolated over the entire bandheads. Interestingly, at the spectral resolution of the X-shooter observations, several individual $J-$components of the rotational ladder are spectrally resolved. This is clearer in the first two bandheads. This fact, together with the sharp blue part of the spectrum to the left of the peak of the bandheads, allows us to accurately constrain $\varv_\mathrm{K}\sin i_\mathrm{disc}$ (see Appendix~\ref{sect:appendix_CO}). Figure~\ref{fig:CO_spectrum} shows the first four bandheads of the CO emission overplotted with our best model. The model parameters giving the best results are $T = 3000^{+500}_{-500}$\,K, $N(\mathrm{CO}) = 1^{+0.2}_{-0.4}\times10^{22}\mathrm{\,cm^{-2}}$, $\varv_\mathrm{K}\sin i_\mathrm{disc} = 25^{+5}_{-10}\mathrm{\,km\,s^{-1}}$, and $\Delta \varv = 10^{+2}_{-3}\mathrm{\,km\,s^{-1}}$ (see Table\,\ref{tab:CO_model_output}). These physical conditions are consistent with previous studies modelling the CO emission around HMYSOs \citep{ilee2013}.

\begin{table}
\caption{Inner disc properties from the LTE CO model.}             
\label{tab:CO_model_output}      
\centering          
\begin{tabular}{l c}    
\hline\hline       
\noalign{\smallskip}
Parameter & Value\\ 
\noalign{\smallskip}
\hline              
\noalign{\smallskip}
Temperature (K) & $3000^{+500}_{-500}$ \\
\noalign{\smallskip}
CO column density $(\mathrm{cm^{-2}})$ & $1^{+0.2}_{-0.4}\times10^{22}$ \\
\noalign{\smallskip}
$\varv_\mathrm{K}\sin i_\mathrm{disc}$ $(\mathrm{km\,s^{-1}})$ & $25^{+5}_{-10}$ \\
\noalign{\smallskip}
$\Delta \varv$ $(\mathrm{km\,s^{-1}})$ & $10^{+2}_{-3}$ \\
\noalign{\smallskip}
\hline                  
\end{tabular}
\end{table}

\begin{figure*}[ht!]
\centering
\includegraphics[width=1.0\textwidth]{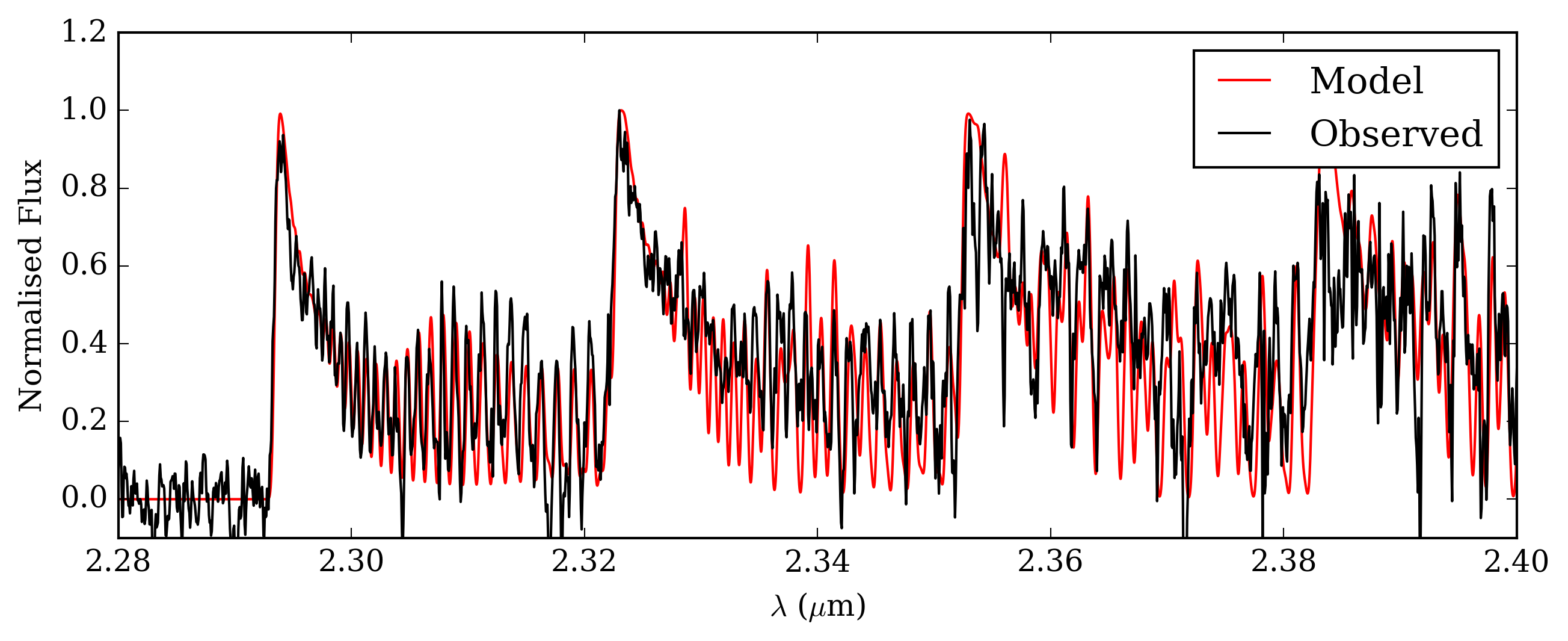}
\caption{Spectrum from VLT/X-shooter of the NIR CO overtone emission (black) and LTE model (red) obtained for $T = 3000$\,K, $N(\mathrm{CO}) = 1\times10^{22}\mathrm{\,cm^{-2}}$, $\varv_\mathrm{K}\sin i_\mathrm{disc} = 25\mathrm{\,km\,s^{-1}}$, and $\Delta \varv = 10\mathrm{\,km\,s^{-1}}$.}
\label{fig:CO_spectrum}
\end{figure*}




\section{Discussion}\label{sect:discussion}

\subsection{CO disc emission reflected in the outflow cavity wall}

Our CO model gives reasonable parameters of a warm $(T=3000\pm500\mathrm{\,K})$ and dense $(N(\mathrm{CO})=1^{+0.2}_{-0.4}\times10^{22}\mathrm{\,cm^{-2}})$ gas consistent with previous results for other sources \citep{ilee2013}. In fact, the high CO column density suggests that this emission possibly originates close to the midplane of the disc. Indeed, there is a factor of $10^{4}$ between the CO column density and the total gas column density, implying a gas column density of $N_\mathrm{total}\sim10^{26}\mathrm{\,cm^{-2}}$.

As suggested by the jet geometry, the circumstellar disc of IRAS\,11101-5829 should be almost edge-on. Such geometry should be very evident in our CO line profiles as the projected Keplerian rotation term ($\varv_\mathrm{K}\sin i_\mathrm{disc}$) is expected to be large in edge-on discs and therefore form a prominent ``shoulder'' to the left of the bandheads; in Figs. 3 and 4 of \citet{ilee2013} the prominent shoulder in a spectrum with large $\varv_\mathrm{K}\sin i_\mathrm{disc}$ is shown (see also Appendix~\ref{sect:appendix_CO} for an explanation of the formation of the shoulder). However, no indication of large $\varv_\mathrm{K}\sin i_\mathrm{disc}$ ($=25\pm5\mathrm{\,km\,s^{-1}}$) is evident in our observations (see Figure~\ref{fig:CO_spectrum_zoom} for a zoom-in in the first bandhead). This term is well constrained because no shoulder is present and the low-$J$ components are spectrally resolved, which allow us to match the expected wavelength. There are then two possible explanations for the low value of $\varv_\mathrm{K}\sin i_\mathrm{disc}$.

\begin{figure}[ht!]
\includegraphics[width=0.485\textwidth,right]{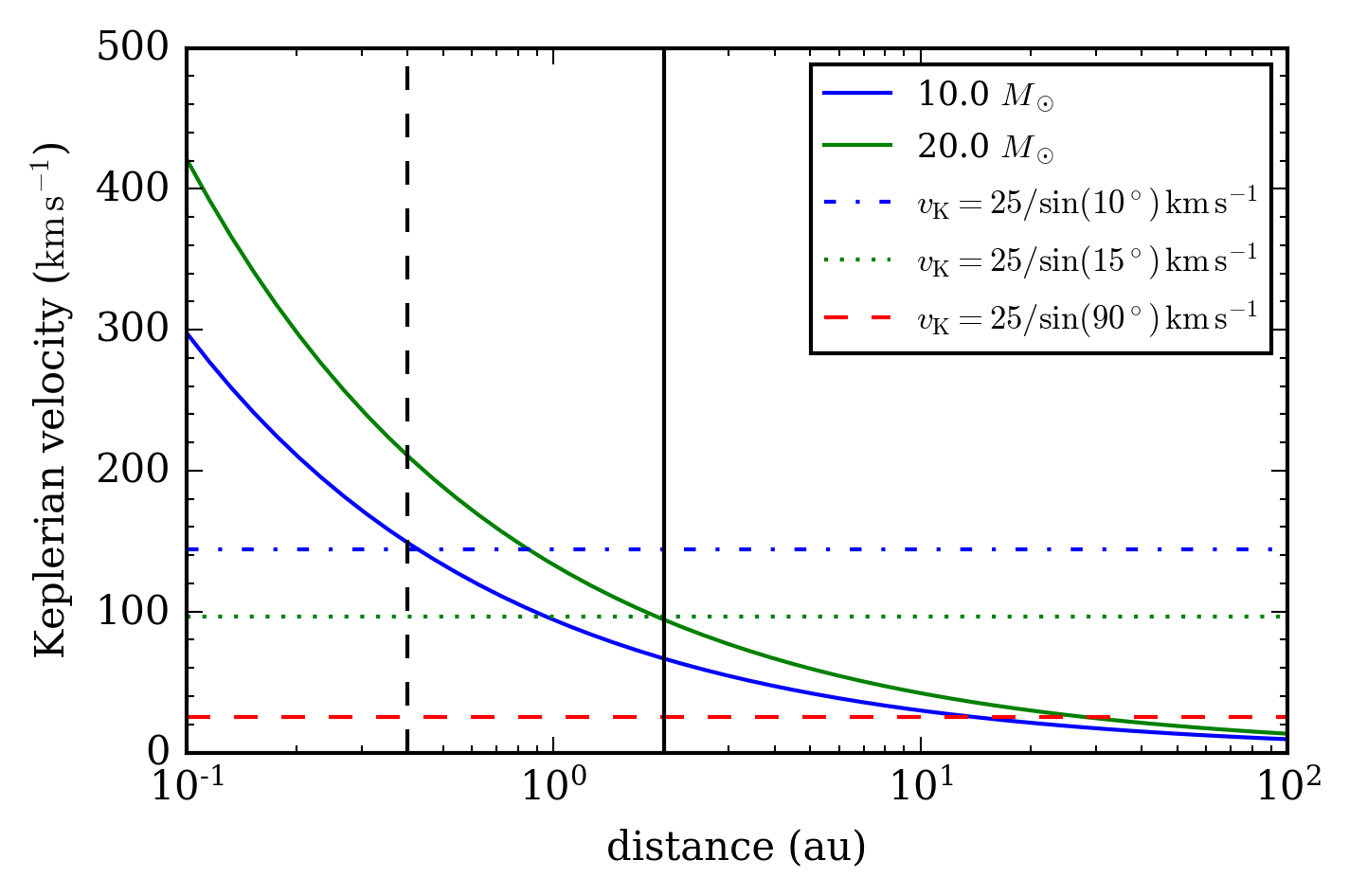}
\includegraphics[width=0.49\textwidth,right]{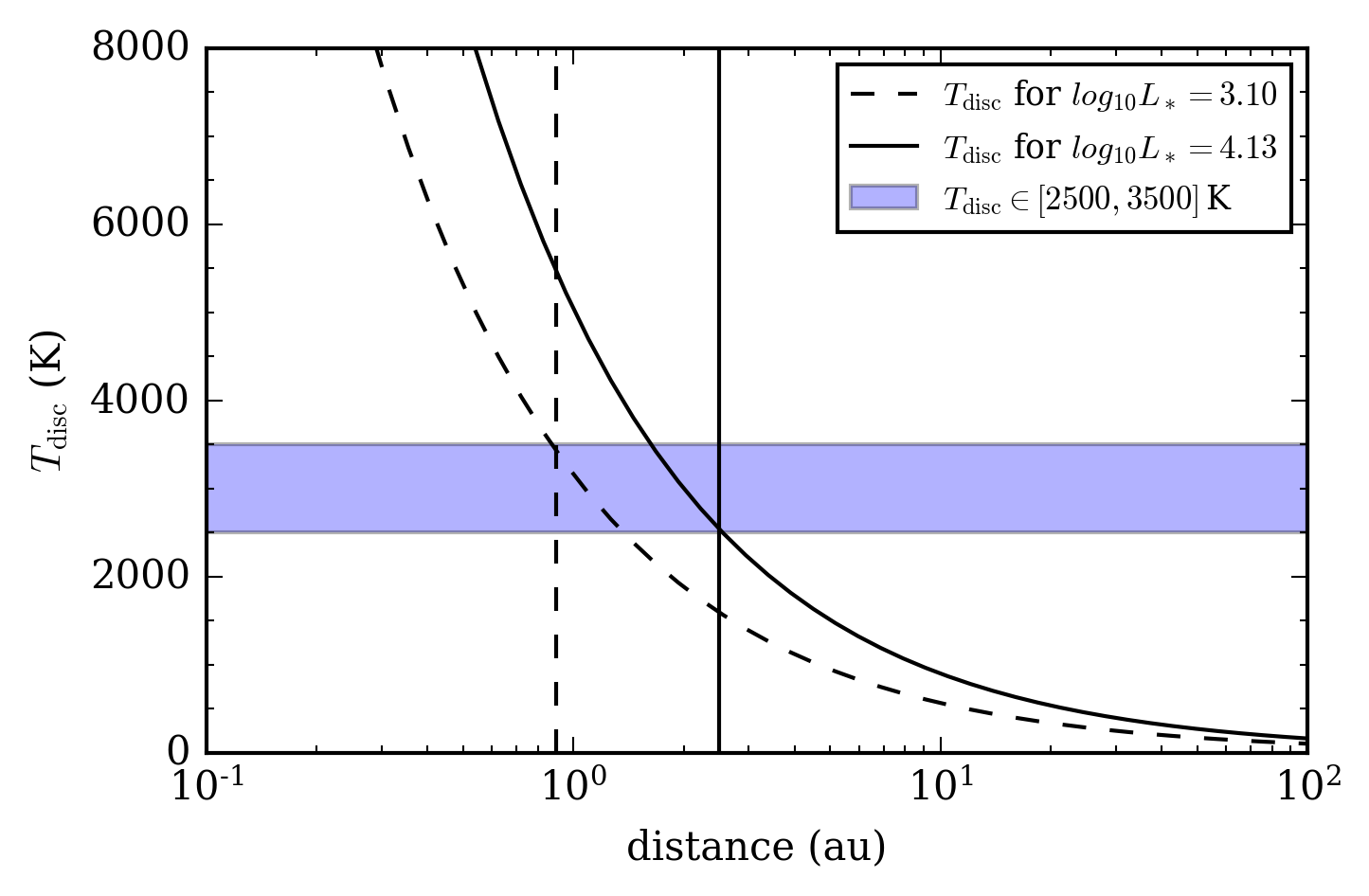}
\caption{\textit{Top panel:} Keplerian rotation vs. distance from the central source for a $10\,M_\odot$ (blue solid curve) and $20\,M_\odot$ (green solid curve) protostar. Dashed lines show the expected Keplerian velocity for $i$ equals 10$\degr$, 15$\degr$, and 90$\degr$ (blue, green, and red dashed line, respectively). Vertical lines at 0.4\,au (dashed) and 2.0\,au (solid) are drawn for reference where the lines intersect with the plausible inclination values.
\textit{Bottom panel:} Disc temperature vs. distance from the central source for $log_{10}L_*=3.10$ (black dashed curve) and $log_{10}L_*=4.13$ (black solid curve). Observed gas temperature range is indicated in purple.
Vertical lines at 0.9\,au (dashed) and 2.5\,au (solid) are drawn for reference where the lines intersect with the plausible values in temperature.}
\label{fig:disc_checks}
\end{figure}

The first possibility is that we are observing the CO emission from an almost edge-on disc geometry. Therefore, $\sin i_\mathrm{disc}\sim1$ (hence $i_\mathrm{disc}\sim90^\circ{}$, i.e. edge-on disc) means that $\varv_\mathrm{K}\sin i_\mathrm{disc}\approx \varv_\mathrm{K}\approx25\mathrm{\,km\,s^{-1}}$. However, this velocity would locate the CO emitting region at more than 10\,au in a disc in Keplerian rotation ($v_\mathrm{K}=(GM_*/d)^{0.5}$, where $G$ is the gravitational constant, $M_*$ the mass of the source, and $d$ the distance from source) around a massive protostar of $10-20\,M_\odot$ (Fig.~\ref{fig:disc_checks} top panel). However, this scenario can be excluded when considering the gas expected temperature at that distance. A simple estimate of the temperature in the midplane of the gaseous disc is given by Eq.~(\ref{eq:T_midplane}) \citep[i.e. Eq.~(15) of][]{dullemond2010}, under the assumption of a dust-free gaseous disc heated by stellar radiation, i.e.

\begin{equation}
    T_\mathrm{disc} = \left(\frac{R_*^{3}f}{3\pi}\right)^{1/4}T_*R_\mathrm{disc}^{-3/4}
    \label{eq:T_midplane}
\end{equation}

\noindent where $R_*,R_\mathrm{disc},T_*, f$ are the radius of the star, the distance from the central source, the temperature of the star, and the factor describing how much of the stellar radiation is longwards of $0.45\,\mu$m, which was taken to be one-half, respectively.

As $L_\mathrm{bol}\sim10^4\,L_\odot$, we consider two $L_*$ values to cover a wide range of possible luminosities. Firstly, we assume $L_*\sim L_\mathrm{bol}$ as an upper limit (Fig.~\ref{fig:disc_checks} bottom panel). Secondly, we assume $L_*\sim10^3\,L_\odot$ as a lower limit considering that accretion luminosity is comparable to the stellar luminosity. Stellar parameters were taken from \citet{pecaut2013} assuming the star on the zero age main sequence (ZAMS). Figure~\ref{fig:disc_checks} bottom panel shows that in both cases the gas temperature at more than 10\,au drops to few hundred Kelvin, which  disagrees with our CO modelling (range of values in $T_\mathrm{disc}$ shown in light blue area). As the stellar radiation might not be the main source of disc heating, we also investigated the heating effect from viscous accretion \citep[see e.g. Eq. (7.2) from][]{hartmann2009}. We find that this effect is not dominant for mass accretion rates equal or below $10^{-4}\,M_\odot\,\mathrm{yr^{-1}}$. Finally, both the large distance and low temperature of the CO emitting region are also in disagreement with previous studies \citep{ilee2013}. Therefore, we can initially exclude the edge-on scenario.

The second possibility is that we are observing the CO emission from an almost face-on disc geometry. Our observations suggest that $i_\mathrm{disc}$ is low ($\sim10-15^\circ{}$, almost face-on disc) since the CO bandhead line profile is characterised by a very sharp or non-existent shoulder (see Figs.~\ref{fig:CO_spectrum} and \ref{fig:CO_spectrum_zoom}). A low value of $i_\mathrm{disc}$ implies that $\varv_\mathrm{K}\sin i_\mathrm{disc}\approx100-150\mathrm{\,km\,s^{-1}}$ (green dotted and blue dash-dotted lines in Fig.~\ref{fig:disc_checks}, respectively), which agrees better with the expected distance of the CO disc emitting region, i.\,e. a few astronomical units. At this distance, the temperature of the disc ($T_\mathrm{disc}$) is consistent with the results obtained in our modelling (see Fig.~\ref{fig:disc_checks}) and with results from literature \citep{ilee2013}. Nevertheless, a face-on disc geometry of the disc strongly disagrees with that of the jet, which is well-constrained.

\begin{figure}[ht!]
\centering
\includegraphics[width=0.45\textwidth]{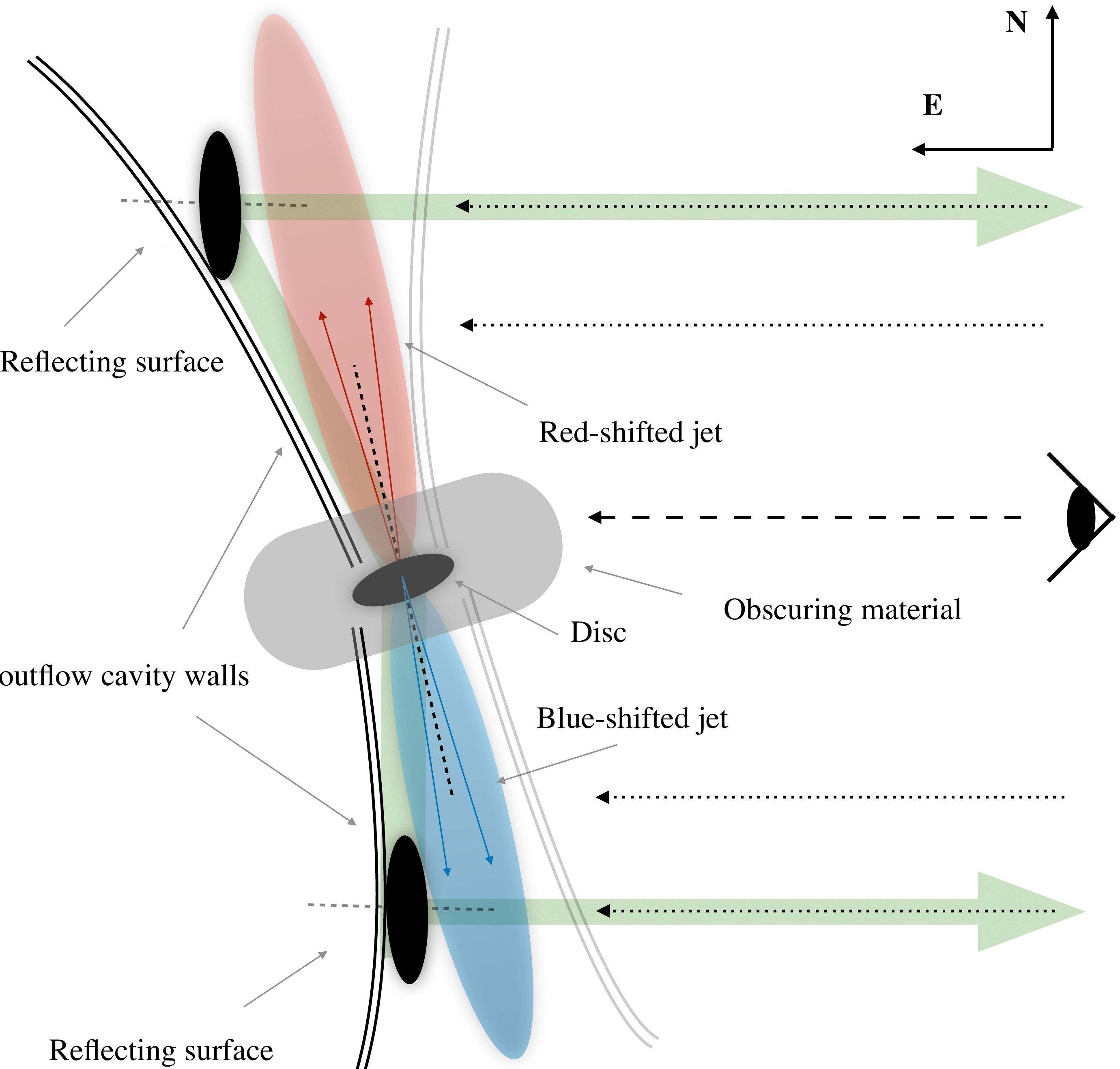}
\caption{Suggested configuration of the first 20\,000 au of the IRAS\,11101-5829 system. There is some obscuring material blocking direct observation of the central source and its immediate environment. The system is reflected in the outflow cavity walls.}
\label{fig:schematic_view}
\end{figure}

We suggest a reflection scenario to reconcile the disc-jet geometries. In the proposed scenario, the disc of IRAS\,11101-5829 is indeed edge-on (in the plane of the sky), but what we are observing is the reflected light in the mirror of the outflow cavity walls, from which the disc is ``seen'' close to face-on. In Figure~\ref{fig:schematic_view} we present our proposed reflection scenario highlighting the main components. Direct observations of the circumstellar disc is blocked by obscuring material, most likely the thick envelope of the system. This prevents us from observing the characteristic shoulder of edge-on discs. The light from the disc is then reflected in the outflow cavity wall and observed at a different angle. The large-scale jet is still visible beyond the envelope and allows us to establish the geometry of the system.

Finally, it is worth mentioning that reflected CO emission has been observed in another HMYSO, W33A, using IFU observations as well \citep{davies2010}. The authors observed extended CO and extracted the spectrum at five representative positions. They were able to observe changes in the CO line profile and discussed the possibility that they were observing the system from different viewing angles. There, at variance with IRAS\,11101-5829, the NIR continuum peak matches the real position of the object.

\subsection{Position of the central source}

Another interesting result from our spectroscopic and kinematic analysis is that the position of the driving source in the NIR FoV might be misinterpreted. It is natural to think that its position coincides with the peak of the NIR continuum emission. However, our analysis shows that the CO emission at this position can be modelled with the same disc conditions as in other parts of the outflow cavities. Therefore, the NIR continuum peak is also seen in scattered light. This is also supported by $JHK$ polarimetric observations  \citep{tamura1997}. These observations reveal high degree of polarisation ($30-75\%$) with a centro-symmetric pattern owing to scattered light (see their Fig.~3). Tamura et al. proposed that the scattered polarisation originates from the nebulosity associated with the outflow cavity walls. The authors also proposed a position of the central source from the centro-symmetric pattern of the polarimetric observations in the $K$-band, although with a large uncertainty given by their low spatial resolution ($\sim1.7^{\prime\prime}$; see Figure~\ref{fig:flux_density_maps}). With our observations we can narrow down the uncertainty on the position of the central engine using kinematic maps (see Fig.~\ref{fig:H2_vel_map}). The driving source of HH\,135/136 should lay in the geometrical centre between the blue- and red-shifted lobes of the jet. Therefore, we suggest a position for the driving source of RA(J2000) = 11h12m18.03s, Dec(J2000) = -58d46m21.4s. We estimate an uncertainty of $0.4^{\prime\prime}$ based on our AO-assisted observations (see Fig.~\ref{fig:H2_vel_map}).

\section{Conclusions}\label{sect:conclusions}

In conclusion, we argue that the reflection scenario might be more common than previously expected in other HMYSOs, preventing us from retrieving the true geometry and ultimately the right parameters. Combining simultaneous studies of jet-disc systems can usher us to discern the geometry and nature of these massive protostars better. Likewise, it is fundamental to observe at both high spatial and spectral resolution at the same time to resolve the immediate environment of the central engine. Future ALMA observations will reveal the true geometry of the circumstellar disc of IRAS 11101-5829. At these wavelengths it could be possible to penetrate the obscuring material hindering direct observations of the system in the optical and NIR, reveal the dusty disc, and set strong constraints in its geometry \citep[see e.g.][]{sanchez-monge2013}. Our main findings are summarised in the following:

\begin{itemize}
    \item We observed, for the first time, NIR CO overtone bandhead emission in IRAS\,11101-5829 indicative of an active accretion disc.
    \item We modelled the CO emission with an LTE model retrieving relatively warm $(T=3000\pm500\mathrm{\,K})$ and dense $(N(\mathrm{CO})=1^{+0.2}_{-0.4}\times10^{22}\mathrm{\,cm^{-2}})$ conditions. The high density indicates that this emission is most likely coming from the midplane of the inner gaseous disc.
    \item We also observed the protostellar jet close to the source in the form of H$_2$. We estimated the geometry of the jet close to the source, which lies in the plane of the sky (consistent with previous results).
    \item Both imaging and spectroscopic analysis indicate that the CO (and the bulk of the Br$\gamma$) emission is reflected in the outflow cavity walls and thus the inner gaseous disc appears to be seen close to face-on.
    \item The NIR continuum emission is seen through scattered light and as a consequence the position of the central source does not coincide with the NIR peak.
\end{itemize}


\begin{acknowledgements}
R.F. acknowledges support from Science Foundation Ireland (grant 13/ERC/12907) and from the School of Physics of University College Dublin. A.C.G. and T.P.R. have received funding from the European Research Council (ERC) under the European Union's Horizon 2020 research and innovation programme (grant agreement No.\ 743029). R.G.L has received funding from the European Union's Horizon 2020 research and innovation programme under the Marie Sk\l{}odowska-Curie Grant (agreement No.\ 706320). M.K. is funded by the Irish Research Council (IRC), grant GOIPG/2016/769 and Science Foundation Ireland (grant 13/ERC/12907). A.N. acknowledges the kind hospitality of the DIAS. We would like to thank the anonymous referee for his/her comments that help to improve the quality of the manuscript. We would like to thank Devaraj Rangaswamy for fruitful conversations on polarimetric observations.
\end{acknowledgements}



\bibliography{phd_bibliography}{}
\bibliographystyle{aa}


\appendix

\section{LTE NIR CO overtone disc model}\label{sect:appendix_CO}

Our CO modelling is based on \citet{kraus2000} and first used in \citet{koutoulaki2019}. We assume that the emission is coming from a single ring undergoing Keplerian rotation in LTE. In this way, the levels are populated following the Boltzmann distribution, i.e.

\begin{equation}
    N_{\upsilon,J}=\frac{N}{Z}(2J+1)\,e^{\frac{-E_{\upsilon,J}}{kT}}
    \label{eq:boltzmann}
,\end{equation}

\noindent where $N$ is the total column density of the CO molecule, $T$ is the temperature, $J$ is the rotational level (we considered 100 levels), $\upsilon$ is the vibrational level, and $k$ is the Boltzmann constant.\ The quantity $E_{\upsilon,J}$ is the energy of the vibrational-rotational transition given by \citep{dunham1932a,dunham1932b}

\begin{equation}
    E_{\upsilon,J} = hc \sum_{k,l}Y_{k,l}\left(\upsilon+\frac{1}{2}\right)^k(J^2+J)^l
    \label{eq:energy_function}
,\end{equation}

\noindent where $h$ is the Planck constant, $c$ the speed of light, and $Y_{k,l}$ are the Dunham coefficients (where $k$ and $l$ are the power of the vibrational and rotational quantum number, respectively), and $Z$ is the partition function given by the product of the vibrational and rotational partition functions as shown in Eq.\,\ref{eq:partition_function}

\begin{equation}
    Z = Z_\upsilon \cdot Z_J = \sum_\upsilon e^{\frac{-E_\upsilon}{kT}} \sum_J (2J+1)\, e^{\frac{-E_J}{kT}}
    \label{eq:partition_function}
.\end{equation}

The intensity of the line is given by

\begin{equation}
    I_\nu = B_\nu(T)\left(1-e^{-\tau_\nu}\right)
,\end{equation}

\noindent where $B_\nu(T)$ is the Planck function and $\tau_\nu$ is the optical depth for a given frequency $\nu$, which can be written as (but it is actually given by Eq.~\ref{eq:total_tau})

\begin{equation}
    \tau_\nu(\upsilon,J;\upsilon',J') = \frac{c^2A_{\upsilon,J;\upsilon',J'}}{8\pi\nu^2}\left(\frac{2J+1}{2J'+1}\frac{N_{\upsilon',J'}}{N_{\upsilon,J}}-1\right)N_{\upsilon,J}
    \label{eq:optical_depth}
.\end{equation}

\noindent We assume that each $J$ component has a Gaussian profile given by

\begin{equation}
    \Phi(\nu) = 
    \frac{1}{\sqrt{2\pi}\frac{\Delta \varv}{c}\nu_0}
    \exp\left[-\left(\frac{\nu-\nu_0}{2\frac{\Delta \varv}{c}\nu_0}\right)^2\right]
    \label{eq:guassian_broadening}
,\end{equation}

\noindent where $\nu_0$ is the rest frequency of the $J$ component. Applying this broadening  a total optical depth is obtained given by

\begin{equation}
    \tau(\nu) = \sum_{J=0}^{J=100}\tau_\nu(\upsilon,J)\Phi(\nu)
    \label{eq:total_tau}
.\end{equation}

The way we select the model velocity grid generates a spectrum with a resolution of $\mathcal{R}_\mathrm{model}\sim0.3\mathrm{\,km\,s^{-1}}$. Once the spectrum is generated and before comparing with the observations, we reduced the resolution to the observed one using a Gaussian convolution (in the case of X-shooter to $\mathcal{R}\sim 7000$, i.e. $\sim43\mathrm{\,km\,s^{-1}}$).

\subsection{Expression $\varv_\mathrm{K}\sin i_\mathrm{disc}$ in the CO modelling}

Assuming that the CO is emitted in a ring in Keplerian rotation, the rotational velocity is

\begin{equation}
    \varv_\mathrm{K} = \sqrt{\frac{GM_*}{r}}
    \label{eq:keplerian_velocity}
,\end{equation}

\noindent where $G$ is the gravitational constant, $M_*$ the protostar mass, and $r$ the disc radius or distance from the central source. For a single element of the ring with azimuthal angle $\theta\in[0,2\pi],$ the projected velocity is given by the expression

\begin{equation}
    \varv_\mathrm{proj} = \cos\theta\,\varv_\mathrm{K}\sin i_\mathrm{disc}
    \label{eq:projected_velocity}
,\end{equation}

\noindent where $i_\mathrm{disc}$ is the inclination of the plane of the disc with respect to the plane of the sky, i.e. $i_\mathrm{disc}=0^\circ{}$ is a face-on disc and $i_\mathrm{disc}=90^\circ{}$ is an edge-on disc. We note that $\varv_\mathrm{proj}$ takes both positive and negative numbers since $\cos\theta\in[-1,1]$. This fact determines the line broadening and is more evident the larger $\varv_\mathrm{K}\sin i_\mathrm{disc}$ gets. Owing to rotation, the frequency is Doppler shifted according to the expression

\begin{equation}
    \nu_\mathrm{K} = \nu_\mathrm{int}\left(1+\frac{\varv_\mathrm{proj}}{c}\right)
    \label{eq:keplerian_frequency}
,\end{equation}

\noindent where $\nu_\mathrm{int}$ is the intrinsic frequency. This is the term responsible of the characteristic blue shoulder created in systems with large values of $\varv_\mathrm{K}\sin i_\mathrm{disc}$ and therefore large $\varv_\mathrm{proj}$.

\begin{figure*}[ht!]
\centering
\includegraphics[width=0.9\textwidth]{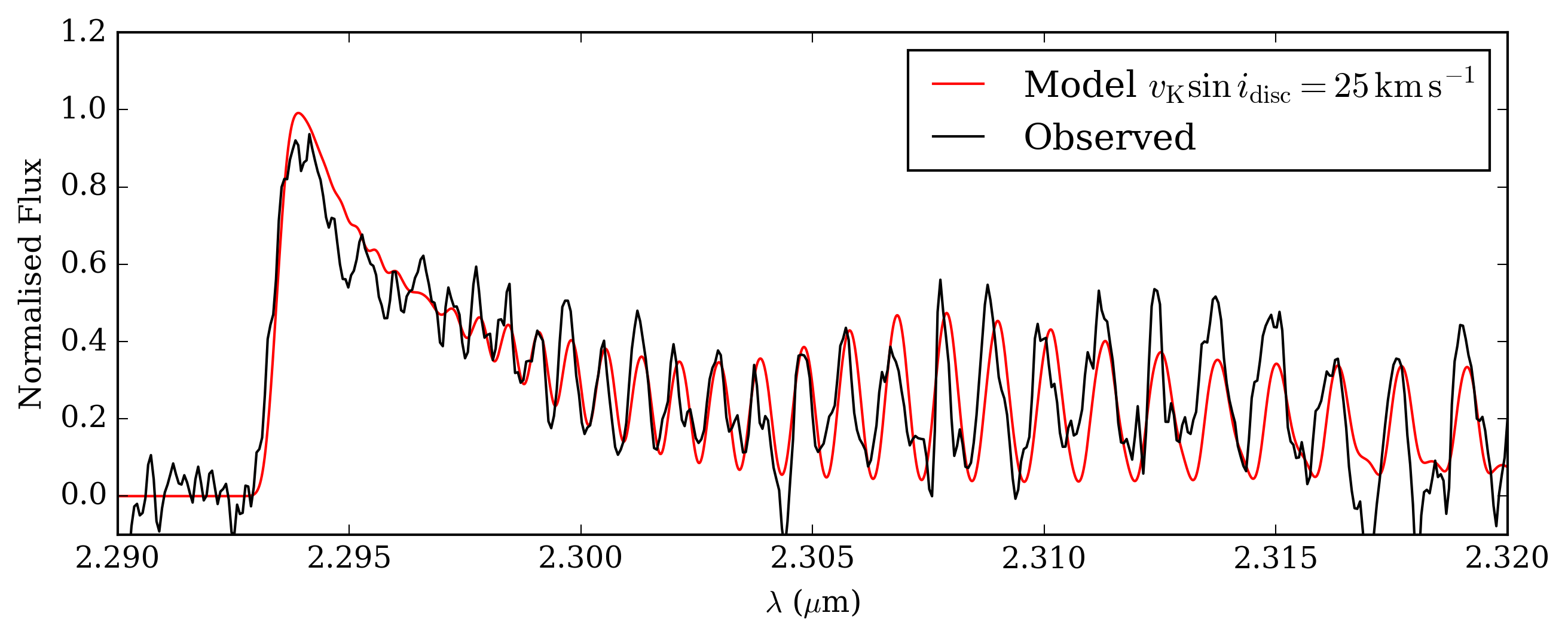}
\includegraphics[width=0.9\textwidth]{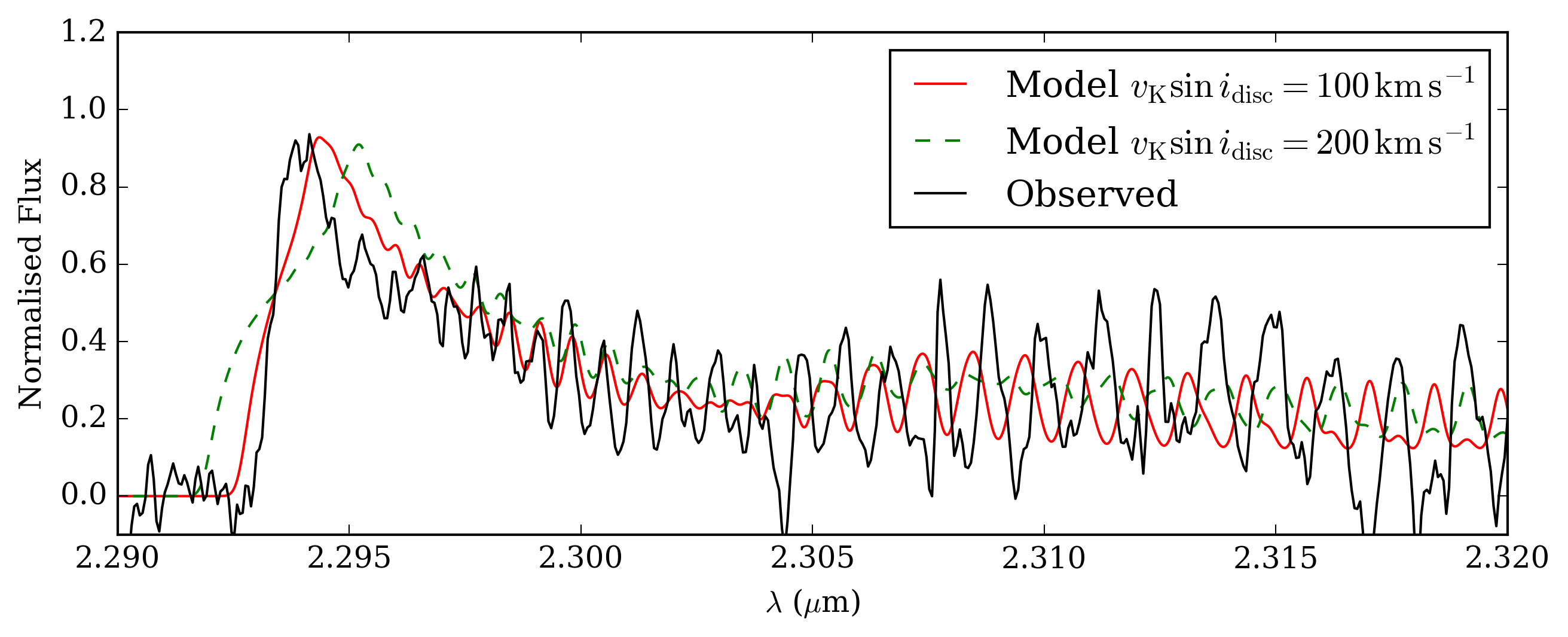}
\caption{\textit{Top panel:} VLT/X-shooter spectrum of the first CO bandhead emission (black) and the LTE model spectrum (red) obtained for $T = 3000$\,K, $N(\mathrm{CO}) = 1\times10^{22}\mathrm{\,cm^{-2}}$, $\varv_\mathrm{K}\sin i_\mathrm{disc} = 25\mathrm{\,km\,s^{-1}}$, and $\Delta \varv = 10\mathrm{\,km\,s^{-1}}$. \textit{Bottom panel:} Same as top panel but with $\varv_\mathrm{K}\sin i_\mathrm{disc} = 100\mathrm{\,km\,s^{-1}}$ (red solid line) and $200\mathrm{\,km\,s^{-1}}$ (green dashed line).}
\label{fig:CO_spectrum_zoom}
\end{figure*}

Even though an accurate and fully constrained fit is not possible with the resolution of our observations $(\mathcal{R}\sim7000)$, important information can still be retrieved. Therefore a visual fit was performed to estimate the best parameters. In particular, the term $\varv_\mathrm{K}\sin i_\mathrm{disc}$ can be very well constraint without the high spectral resolution $(\mathcal{R}\gtrsim30\,000)$ needed for an accurate and proper fit \citep[see e.\,g.][]{ilee2013,ilee2014}. Fixing a set of $T$, $N(\mathrm{CO})$, and $\Delta \varv$, a small variation in $\varv_\mathrm{K}\sin i_\mathrm{disc}$ completely changes the shape of the CO emission. To prove this, we ran our code for the values that best reproduce our observations (Fig.~\ref{fig:CO_spectrum_zoom}, top panel) and for the same conditions, adopting values of $\varv_\mathrm{K}\sin i_\mathrm{disc}$, which are large but still reasonable, $\varv_\mathrm{K}\sin i_\mathrm{disc}=100-200\mathrm{\,km\,s^{-1}}$ \citep[see][for observations reproduced with large values of $\varv_\mathrm{K}\sin i_\mathrm{disc}$]{ilee2013}. We can see that with a large value of $\varv_\mathrm{K}\sin i_\mathrm{disc}$, the characteristic blue shoulder appears (see above). In addition to that, this term produces a ``shift'' of the spectrum to the left or to the right, and therefore affects all the $J-$components. In Figure~\ref{fig:CO_spectrum_zoom} bottom panel, it is clear that, by applying a large value of $\varv_\mathrm{K}\sin i_\mathrm{disc}$, the low-$J$ components and high-$J$ components are missed in the creation of the blue shoulder. In conclusion, using the shape of the left part of the peak of the bandhead together with the shape of the low-$J$ components in the observations, we are able to estimate $\varv_\mathrm{K}\sin i_\mathrm{disc}$ in the model. In this way, we estimated the uncertainty in our parameters, by fixing all variables but one and varying it until the shape of the CO was significantly different. Observations at higher spectral resolution are needed to accurately and properly fit our observations. Model output parameters are given in Table~\ref{tab:CO_model_output}.

\begin{figure*}[ht!]
\centering
\includegraphics[width=0.9\textwidth]{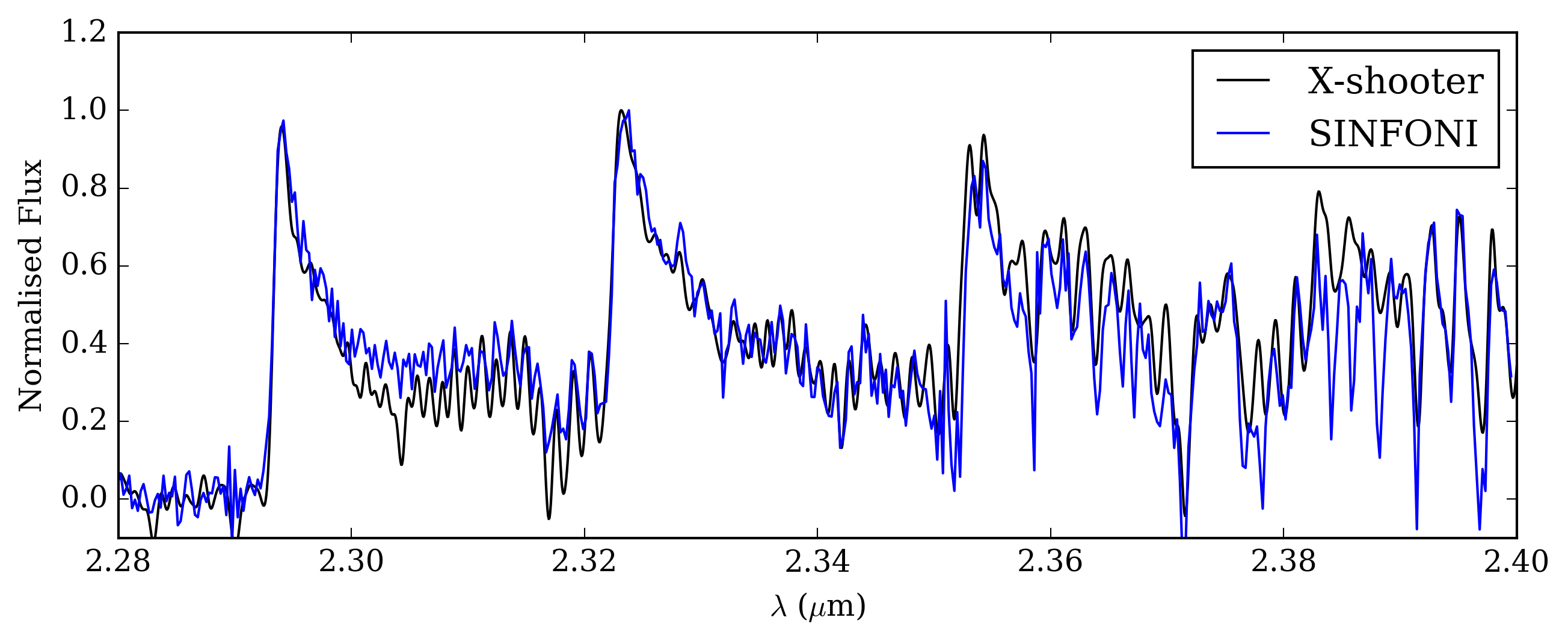}
\caption{VLT/X-shooter spectrum (black) in the CO emission range plotted over the VLT/SINFONI spectrum (blue). The resolution of the X-shooter spectrum was downgraded to 4000 to match that of SINFONI.}
\label{fig:CO_spectrum_xs_sin}
\end{figure*}

\end{document}